\documentclass[
reprint,amssymb,
amsmath,nobibnotes,siunitx,aps,prm,
floatfixaps,citeautoscript,showkeys,
floatfix,longbibliography,balance,
superscriptaddress,
bibnotes,prm,
]{revtex4-2}
\usepackage[table,dvipsnames]{xcolor}
\usepackage[utf8]{inputenc}
\usepackage[T1]{fontenc}
\usepackage{graphicx}
\graphicspath{{./plots/}}
\usepackage{grffile}
\usepackage{longtable}
\usepackage{threeparttable}
\usepackage{multirow}
\usepackage{booktabs}
\usepackage{wrapfig}
\usepackage{rotating}
\usepackage[normalem]{ulem}
\usepackage{amsmath}
\usepackage{textcomp}
\usepackage{amssymb}
\usepackage{times}
\usepackage{capt-of}
\usepackage{hyperref}
\usepackage{lipsum}
\usepackage[]{hyperref}
\hypersetup{colorlinks,pdfborder={0,0,0},linkbordercolor={white},citecolor=blue,filecolor=blue,linkcolor=blue,urlcolor=blue}
\usepackage{pbox}
\usepackage{diagbox}
\usepackage{makecell}
\usepackage[UKenglish,USenglish,english,american]{babel}
\usepackage[]{upgreek}

\date{\today}
\begin{document}
\newcolumntype{x}[1]{>{\centering\arraybackslash\hspace{0pt}}p{#1}}

\bibliographystyle{apsrev4-2}
\pagenumbering{gobble}

\title{\huge Classical and Machine Learning Interatomic Potentials for BCC Vanadium}
\author{Rui Wang}
\affiliation{Department of Materials Science and Engineering, City University of Hong Kong, Hong Kong, China}
\author{Xiaoxiao Ma}
\affiliation{Department of Materials Science and Engineering, City University of Hong Kong, Hong Kong, China}
\author{Linfeng Zhang}
\affiliation{AI for Science Institute, Beijing, China}
\author{Han Wang}
\affiliation{Laboratory of Computational Physics, Institute of Applied Physics and Computational Mathematics, Beijing, China}
\author{David J. Srolovitz}
\affiliation{Department of Mechanical Engineering, The University of Hong Kong, Hong Kong, China}
\author{Tongqi Wen}
\email[Corresponding author: ]{tongqwen@hku.hk}
\affiliation{Department of Mechanical Engineering, The University of Hong Kong, Hong Kong, China}
\author{Zhaoxuan Wu}
\email[Corresponding author: ]{zhaoxuwu@cityu.edu.hk}
\affiliation{Department of Materials Science and Engineering, City University of Hong Kong, Hong Kong, China}
\affiliation{Hong Kong Institute for Advanced Study, City University of Hong Kong, Hong Kong, China}
\date{\today}

\begin{abstract}

  BCC transition metals (TMs) exhibit complex temperature and strain-rate dependent plastic deformation behaviour controlled by individual crystal lattice defects.  Classical empirical and semi-empirical interatomic potentials have limited capability in modelling defect properties such as the screw dislocation core structures and Peierls barriers in the BCC structure.  Machine learning (ML) potentials, trained on DFT-based datasets, have shown some successes in reproducing dislocation core properties.  However, in group VB TMs, the most widely-used DFT functionals produce erroneous shear moduli \(C_{44}\) which are undesirably transferred to machine-learning interatomic potentials, leaving current ML approaches unsuitable for this important class of metals and alloys.  Here, we develop two interatomic potentials for BCC vanadium (V) based on (i) an extension of the partial electron density and screening parameter in the classical semi-empirical modified embedded-atom method (XMEAM-V) and (ii) a recent hybrid descriptor in the ML Deep Potential framework (DP-HYB-V).  We describe distinct features in these two disparate approaches, including their dataset generation, training procedure, weakness and strength in modelling lattice and defect properties in BCC V.  Both XMEAM-V and DP-HYB-V reproduce a broad range of defect properties (vacancy, self-interstitials, surface, dislocation) relevant to plastic deformation and fracture.  In particular, XMEAM-V reproduces nearly all mechanical and thermodynamic properties at DFT accuracies and with \(C_{44}\) near experimental value.  XMEAM-V also naturally exhibits the anomalous slip at 77 K widely observed in group VB and VIB TMs and outperforms all existing, publically available interatomic potentials for V. The XMEAM thus provides a practical path to developing accurate and efficient interatomic potentials for nonmagnetic BCC TMs and possibly multi-principal element TM alloys.

\end{abstract}

\maketitle
\pagenumbering{arabic}

\section{Introduction}

BCC transition metals (TMs) and alloys are an important class of structural materials for high temperature, high strength, radiation or corrosion-resistance applications~\cite{rieth_2013_jnm}.  Their plastic deformation and fracture behaviour are critical for flaw tolerance and structure integrity~\cite{felkins_1998_jom}.  Controlling plasticity and fracture is thus essential but challenging, since both are governed by crystal lattice defect generation, interaction and evolution spanning multiple time and length scales~\cite{elatwani_2018_am}.  For example, point defects such as self-interstitials are generated at individual atomic sites within a few femtoseconds during ion irradiation but their diffusion can occur across multiple grains and over the entire service life of the components~\cite{elatwani_2018_am}.  The challenge is further amplified in the BCC TM family with plenty of surprises seen in experiments~\cite{christian_1983_mta}.  It is now well-established that BCC TMs exhibit strong temperature and strain-rate dependent yield~\cite{felkins_1998_jom,brunner_2000_ml}, non-Schmid/anomalous slip~\cite{seeger_2002_pssa}, parabolic hardening~\cite{brunner_2000_ml} and planar slip at low temperatures~\cite{butler_2018_ijrmhm}, as well as a ductile-to-brittle transition (DBT~\cite{felkins_1998_jom,gumbsch_1998_science}) below some characteristic temperatures, all of which are distinctly different from the behaviour in FCC metals.

At low temperatures, dislocation plasticity is governed by the glide of individual \(\langle 111 \rangle/2\) screw dislocations, which have a high lattice friction (or Peierls barrier, e.g., in W~\cite{cereceda_2013_jpcm}) and require thermally activated, double-kink nucleation and propagation~\cite{zhao_2018_msmse}.  The controlling mechanism for the DBT is, arguably related to the dislocation mobility and nucleation at crack tips, which in turn are influenced by the high lattice friction of the screw/mixed dislocations in the crack near field~\cite{gumbsch_1998_science,romaner_2021_am}.  Nevertheless, the sharp transition temperature suggests that DBT may be related to a switch of defect properties or deformation mechanism at the DBT temperature.   Among the individual BCC elements, subtle differences have also been reported in their defect properties and deformation behaviour, including ground-state self-interstitial structures~\cite{ma_2019_prm}, solute hardening/softening response~\cite{trinkle_2005_science,ghafarollahi_2021_am}, activation of twinning, and dominant dislocation slip planes~\cite{weinberger_2013_imr}.

The complexity in the BCC TMs has root in the partially-filled \(d\)-bands, non-close-packed crystal structure and associated defect properties.  First-principles density functional theory (DFT) calculations have thus been employed extensively to provide quantitative information on fundamental defect properties, such as the generalized stacking fault energy, dislocation core structure and Peierls barrier.  In particular, DFT calculations have unequivocally determined the non-degenerate (ND) core structure of the \(\langle 111 \rangle/2\) screw dislocations (e.g., Ta~\cite{beigi_2000_prl,woodward_2002_prl}, Mo~\cite{beigi_2000_prl,woodward_2002_prl,frederiksen_2003_pm}, W~\cite{romaner_2010_prl}, and Fe~\cite{frederiksen_2003_pm,ventelon_2007_jcamd,romaner_2014_pml}) and its 2D Peierls potential for all 7 BCC TMs~\cite{dezerald_2014_prb}.  These DFT calculations played an instrumental role in advancing fundamental plasticity theory of BCC structure materials.  However, DFT calculations are computationally expensive and typically limited to a few hundred atoms (or a few thousand valence electrons) and several hundred time steps/femtoseconds.  They can not provide the necessary length and time scales required to study dislocation interactions, evolutions and their temperature-dependent behaviour.

Classical interatomic potentials (e.g., embedded-atom method~\cite{daw_1984_prb}, modified embedded-atom method~\cite{baskes_1987_prl,baskes_1989_prb}, bond-order~\cite{pettifor_2004_pms,pastewka_2012_mrsb,drautz_2015_msmse}, etc.) have been developed to approximate interatomic interactions using empirical but more efficient functions since at least early 1980s.  With these classical interatomic potentials, molecular dynamics/statics simulations were performed to study defect dynamics in statistically meaningful ensembles at much larger scales, reaching \(\sim\)100 nanometers and \(\sim\)10 nanoseconds.  For BCC TMs, interatomic potential-based studies have thus been actively pursued over the last few decades, with well over a dozen potentials developed for Fe alone~\cite{wang_2021}.  Those potentials, particularly the most widely used EAM and MEAM ones, had mixed receptions; their physical relevance is often scrutinised against DFT calculations.  In particular, nearly 1/3 out of 72 interatomic potentials examined exhibit a degenerate (D) core structure, in stark contrast to the ND core predicted by DFT~\cite{wang_2021}.  The D-core is thus often considered as an artefact of such interatomic potentials.  Among those possessing the correct ND core, about half have the single-hump Peierls energy profile as that predicted by DFT, and only a few of them have quantitative accuracy in the Peierls barrier (e.g., Fe~\cite{dragoni_2018_prm}).

The deficiencies are well recognized in interatomic potentials for BCC TMs.  Emerging machine-learning (ML) interatomic potentials have been developed to address some of these issues. ML potentials generally use extensible functions (such as neural networks~\cite{blank_1995_jcp,behler_2007_prl,schutt_2018_jcp}, gaussian approximation~\cite{bartok_2010_prl}, rotation-invariant linear model~\cite{seko_2019_prb}) to map atomic environments to total energies, forces and sometimes the virial stresses of a large set of atomic configurations computed by DFT calculations.  For example, many ML potentials have been fit and reported, including Gaussian approximation potentials (GAP~\cite{bartok_2010_prl,bartok_2013_prb,gap_www}) for Fe~\cite{dragoni_2018_prm}, V/Nb/Ta/W/Mo~\cite{byggmastar_2020_prm}, moment-tensor potentials (MTP~\cite{shapeev_2016_mms}) for Fe~\cite{novikov_2022_npjcm}, spectral neighbor analysis potential (SNAP) for NbMoTaW~\cite{li_2020_npjcm}, and deep potential (DP~\cite{han_2018_ccp,zhang_2018_prl}) for W~\cite{wang_2022_nf}.  These ML potentials have demonstrated good accuracy and transferability with respect to DFT calculations.  For example, the GAP-Fe~\cite{dragoni_2018_prm}, ANN-Fe~\cite{mori_2020_prm} and DP-W~\cite{wang_2022_nf} are shown to exhibit the ND core and Peierls barrier in quantitative agreement with DFT predictions for the first time.

\begin{figure}[!htbp]
  \centering
  \includegraphics[width=0.5\textwidth]{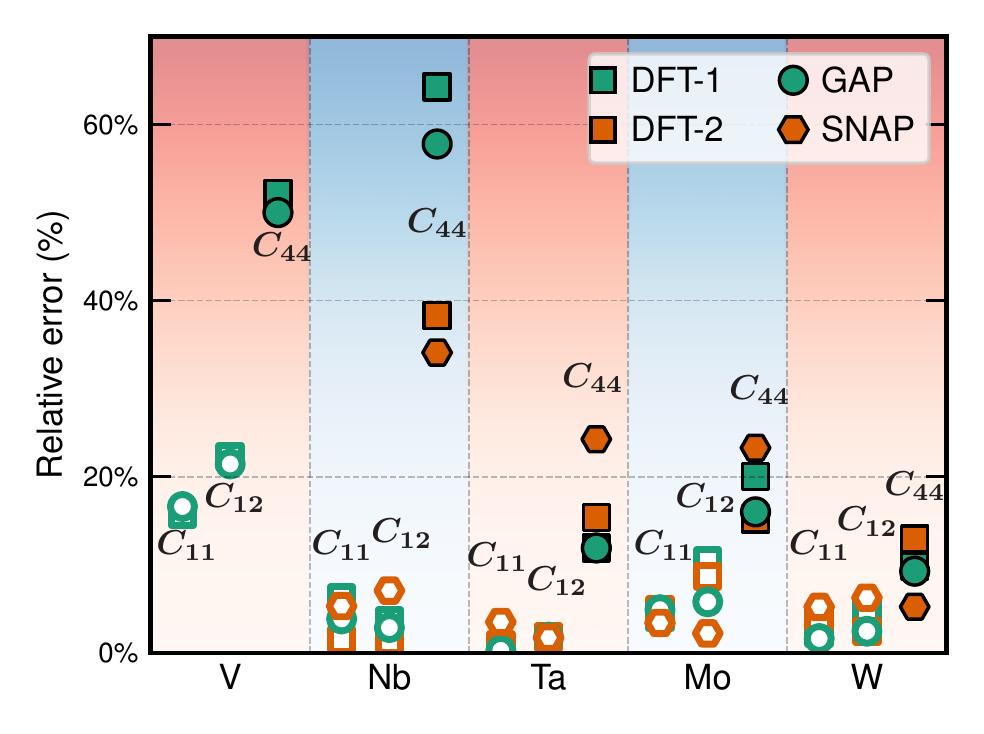}
  \caption{\label{fig:elastic_c_tm} Relative errors of elastic constants of BCC structures predicted by DFT and machine learning interatomic potentials with respect to the experimental values. The errors \(\delta C = \lvert C^\text{model}_{ij} / C^\text{exp}_{ij} -1 \rvert \) are calculated based on values of DFT-1~\cite{byggmastar_2020_prm} and GAP potentials~\cite{byggmastar_2020_prm} for V/Nb/Ta/Mo/W, DFT-2~\cite{li_2020_npjcm} and SNAP potentials~\cite{li_2020_npjcm} for Nb/Ta/Mo/W (see Table~\ref{tab:bcctm_elas_exp_dft_pot}). The machine learning potentials faithfully reproduce the erroneous elastic constants from DFT. The empty symbols are \(C_{11}\) and \(C_{12}\), and the filled symbols are \(C_{44}\).}
\end{figure}

The successes of the ML potentials are remarkable in resolving long-standing issues of modeling dislocations in BCC TMs.  Since ML potentials are fit to atomistic datasets computed by DFT, they are considered to be robust and reliable, provided that current exchange-correlation functionals are exact and can predict material properties accurately.  This is often an implicit and valid assumption in material modelings, as demonstrated in previous DFT calculations~\cite{csonka_2009_prb,lejaeghere_2016_science}.  However, most widely-used DFT methods exhibit deficiencies in predicting some fundamental properties of BCC TMs.  For Group VB TMs (V and Nb), DFT with generalized gradient approximation (GGA)-Perdew–Burke–Ernzerhof (PBE~\cite{perdew_1996_prl}) and other functionals severely underestimates the shear modulus \(C_{44}\) of V and Nb~\cite{wang_2020_jcp} (on the order of 50\% from their respective experimental values, see Fig.~\ref{fig:elastic_c_tm}).  Not surprisingly, ML potentials apparently inherit this deficiency (Fig.~\ref{fig:elastic_c_tm}).  On the other hand, many EAM/MEAM potentials~\cite{lee_2001_prb}, ML GAP-W~\cite{byggmastar_2020_prm} and DP-W~\cite{wang_2022_nf} have accurate \(C_{44}\).  Reproducing the elastic constants of BCC structures thus does not seem to be a gruelling task.  The shortcomings of current ML potentials do not arise from their energy function formalism or learning strategy; these ML potentials faithfully learnt all information produced by DFT.

For BCC V in particular, at least 7 interatomic potentials have been developed and made publically available.  Table~\ref{tab:ex_pots} shows a brief survey of their properties.  Two classical potentials (EAM3~\cite{mendelev_2007_prb} and MEAM2~\cite{maisel_2017_prm}) and the GAP potential~\cite{byggmastar_2020_prm} have the ND screw dislocation core structure as predicted by DFT, while the rest have the D/unstable screw core.  EAM3 and MEAM2 have a double-hump Peierls potential profile, contradicting DFT predictions~\cite{dezerald_2014_prb,weinberger_2013_prb}. Only the recent GAP-V~\cite{byggmastar_2020_prm} has a single-hump Peierls potential profile, but yields a Peierls barrier \(\Delta E_{\text{PB}}\) = 66.9 meV/\(b\), \(\sim\)2.6 times the DFT values (24.4~\cite{weinberger_2013_prb}-25.7~\cite{dezerald_2014_prb} meV/\(b\)) and about 80\% of that of DFT-W~\cite{marinica_2013_jpcm,dezerald_2014_prb}.  GAP-V thus underestimates \(C_{44}\) by 50\% and very likely overestimated \(\Delta E_\text{PB}\).  Considering these issues,  all current interatomic potentials have limited capability for modelling crystal lattice defects in BCC V.

\begin{table*}[!htbp]
  \caption{\label{tab:ex_pots} Comparisons of key properties obtained from extant interatomic potentials, DFT calculations and experiment measurements for V. The properties include the BCC lattice parameter \(a\) (\AA), cohesive energy \(E_\text{c}\) (eV/atom), elastic constant \(C_{44}\) (GPa), FCC-BCC structural energy difference \(\Delta E_{\text{FCC-BCC}}\) (eV/atom), the \(\langle 111 \rangle/2\) screw core structure, Peierls energy profile and Peierls barrier \(\Delta E_{\text{PB}}\) (meV/\(b\)). The Peierls energy profile and barrier are only calculated for potentials with the non-degenerate (ND) core structure.}
  \centering
  \begin{threeparttable}
  \begin{tabular}{x{2.0cm} x{1.5cm} x{1.5cm} x{1.5cm} x{1.5cm} x{1.8cm} x{1.8cm} x{1.8cm} x{2.5cm}}
    \toprule
    Property & EAM1~\cite{han_2003_jap} & EAM2~\cite{derlet_2007_prb} & EAM3~\cite{mendelev_2007_prb} & EAM4~\cite{olsson_2009_cms} & MEAM1~\cite{lee_2001_prb} & MEAM2~\cite{maisel_2017_prm} & GAP-V~\cite{byggmastar_2020_prm} & DFT/\textbf{Exp.} \\
    \hline
    \(a\) & 3.00  & 3.04  & 3.03  & 3.03  & 3.03  & 3.00  & 3.00  & 3.00~\tnote{a} /\textbf{3.03}~\cite{kittel_2019_issp}  \\
    \(E_\text{c}\)    & -5.29 & -5.31 & -5.02 & -5.31 & -5.30 & -5.30 & -5.38 & -5.38~\tnote{a} /-\textbf{5.31}~\cite{kittel_2019_issp} \\
     \(C_{44}\)  &  32550  & 43.5 & 42.0 & 46.0 & 46.0 & 50.1 & 23.7 & 23.6~\tnote{a} /\textbf{46.0}~\cite{simmons_1971_mit}\\
    \(\Delta E_{\text{FCC-BCC}}\)  & 0.157 & 0.123 & 0.214 & 0.138 & 0.084 & 0.244 & 0.242 & 0.243~\tnote{a} \\
    Core structure~\tnote{b}      &  us    & D     & ND     & D     & D     & ND     & ND     & ND ~\cite{weinberger_2013_prb,dezerald_2014_prb}  \\
    PE profile~\tnote{c} & -    & -     & DH     & -     & -     & DH     &  SH   & SH ~\cite{weinberger_2013_prb,dezerald_2014_prb} \\
    \(\Delta E_{\text{PB}}\)  & - & -     & -      & -     & -     & -      &  66.9     & 24.4~\cite{weinberger_2013_prb}, 25.7~\cite{dezerald_2014_prb} \\
    \bottomrule
  \end{tabular}

\begin{tablenotes}
\item[a] DFT in this work.
\item[b] us: the core structure is unstable; D: Degenerate core structure; ND: Non-degenerate core structure.
\item[c] DH for double hump and SH for single hump in the Peierls energy profile.
\end{tablenotes}

\end{threeparttable}
\end{table*}

To address the above problems,  at least two approaches can be attempted and may lead to more accurate interatomic potentials suitable for modelling general plastic and fracture phenomena in BCC V.  One is to increase the fitting flexibility of the classical MEAM by expanding its energy functions, while the other is to train ML potentials with corrected datasets and incorporate core structure information from DFT.  Since the spline-MEAM potentials for Mo~\cite{park_2012_prb} and Nb~\cite{yang_2019_cms} give accurate screw dislocation core structures and Peierls barrier profiles, it seems feasible to use spline-MEAM for BCC TMs.  The analytical functions of the electron density terms can also be extended to higher order terms or use spline-based forms, similar to the multi-state MEAM~\cite{gibson_2017_msmse}.  Separately, a new three-body embedding descriptor has recently been hybridized to the DP framework (DP-HYB~\cite{wang_2022_nf}).  Based on the DP-HYB, a new potential for W has been developed and shown to yield accurate properties (point defects, core structure and Peierls barrier) relavant to its mechanical behaviour.

In this work, we explore both approaches to examine their respective strengths and weaknesses in developing interatomic potentials for an important class of materials and in particular BCC V.  In the classical potential path, we extend the original MEAM formulation by including additional angular electron density terms and use different screening parameters for the embedding function and pair interaction function, similar to the multi-state MEAM~\cite{gibson_2017_msmse}.  We denote this extension of MEAM as XMEAM.  In the XMEAM, we preserve the analytical functions of all electron density terms and the embedding energy function,  which retains the physical interpretation of the classical MEAM.  The XMEAM is shown to give more flexibility compared to MEAM in reproducing many material properties such as the energies of BCC and FCC structures.  For the ML approach, we use the latest DP-HYB due to its enhanced representation and generalization properties~\cite{wang_2022_nf}.  We provide detailed analysis of the XMEAM and DP-HYB in the fitting procedure, accuracies on individual properties and computational efficiency.  Both the XMEAM and DP-HYB yield interatomic potentials for V (XMEAM-V and DP-HYB-V) significantly more accurate than all existing ones.  Nevertheless, DP-HYB-V inherits some of the deficiencies of current DFT calculations (discussed above). Further optimization may be possible via the use of higher order or otherwise enhanced DFT to produce the DP training dataset.  On the other hand, XMEAM-V reproduces an extensive range of properties, making it the better choice of interatomic potential for modelling plastic and fracture behaviour of V at present.  XMEAM-V is then applied to study lattice defect properties and reveals several unique features of dislocation behaviour in V for the first time.  The simulation results are also compared with previous experimental studies.  Like the DP approach, XMEAM is general and can be applied to other nonmagnetic BCC TMs as well.

In the following, we first introduce the general methods and simulation cells used in the DFT calculations and molecular dynamics (MD) simulations, followed by the details on the development of XMEAM-V and DP-HYB-V, as well as the calculation models for individual defects.  In Section~\ref{sec:results}, we present a comprehensive comparison between the resulting classical XMEAM-V and ML potentials DP-HYB-V (and GAP-V) on their thermodynamic and mechanical properties, including point defects, dislocations and their finite-temperature behaviour.  We also perform a quantitative benchmark on the computational speed of these three interatomic potentials. Section~\ref{sec:discussion} discusses the strengths and weaknesses of the classical and ML approaches based on the current results.  We particularly focus on the broad implications on developing accurate interatomic potentials for crystal defects of structure materials.  Section~\ref{sec:conclusion} summarizes the key conclusions and provides an outlook for future works related to interatomic potentials for the BCC TM family.

\section{Methodology and Computational Models}
\label{sec:method}
We first describe the general methods and parameter settings employed in the current work.  These parameters apply to all the calculations unless otherwise mentioned in the respective models.

\subsection{DFT calculations}
\label{sec:method_dft}
The DFT calculations are performed using the Vienna \textit{Ab initio} Simulation Package (VASP~\cite{kresse_1996_cms,kresse_1996_prb,kresse_1999_prb}).  We employ the generalized gradient approximation (GGA) with the Perdew–Burke–Ernzerhof (PBE~\cite{perdew_1996_prl}) exchange-correlation functional.  The outer 13 electrons (\(3s^23p^63d^44s^1\)) of V are treated as valence electrons and the rest as core electrons replaced by the projector-augmented-wave (PAW~\cite{blochl_1994_prb}) pseudopotential (V\_sv).  The plane-wave expansion cutoff energy is set to 650 eV.  We use the Monkhorst–Pack Mesh method~\cite{monkhorst_1976_prb} to sample the Brillouin zone with a \(k\)-points grid spacing of 0.1 {\AA}$^{-1}$. The first-order Methfessel–Paxton method~\cite{methfessel_1989_prb} with a smearing width of 0.22 eV is used for integration in the Brilliouin zone.  During atomic structure optimization, convergence is assumed when the energy variation between two electronic self-consistent steps is below 10$^{-6}$ eV, and all forces after ionic steps are below 0.01 eV/\AA.

The generalized stacking fault \(\gamma\)-lines are calculated for the \(\{110\}\), \(\{112\}\), and \(\{123\}\) planes of the BCC structure.  We use the slab-vacuum supercell and the standard method~\cite{vitek_1968_pm} where atoms are only allowed to move in the direction perpendicular to the slip plane.  In all the cases, the vacuum layer thickness is \(\sim\)20 {\AA}.  The supercells contain 12, 20, and 20 atom layers for the \(\{110\}\), \(\{112\}\), and \(\{123\}\) planes, respectively.  In the equation of state (EOS) calculations, spin-polarized DFT is used to include the influence of magnetic moment on the total energy at large atom separations.

DFT calculations are performed to determine the formation energy of the monvacancy and self-interstitials in BCC V (Fig.~\ref{fig:self_interstitial}). For these point defects, we use a supercell of \(4\times 4\times 4\) cubic unit cells with 128 atoms and the \(k\)-points grid spacing is 0.2 {\AA}$^{-1}$.  The monovacancy/self-interstitials are created by removing/inserting an atom at appropriate positions.  The defect structures are optimised using two methods: (i) fixed supercell (FSC) where supercell vectors are fixed at ideal lattice values and (ii) optimised supercell (OSC) where supercell vectors are optimised to achieve minimum stresses.  Atoms are free to move in both methods.  The former is commonly employed in previous studies~\cite{ma_2019_3_prm}, while the latter allows additional affine deformation and should give a lower formation energy.  Neither of the two methods reproduces the conditions expected in bulk materials, but they should yield consistent results for sufficiently large supercells.

\subsection{Molecular dynamics/statics simulations}
\label{sec:method_md}
All molecular dynamics and statics calculations are performed using the Large-scale Atomic/Molecular Massively Parallel Simulator (LAMMPS~\cite{thompson_2022_cpc}).  Atomic structure optimization is performed with the conjugate gradient method.  In the calculations of surface energy, point defects and \(\gamma\)-surfaces, convergence is assumed when forces on all atoms drop below \(10^{-12} \text{ eV/\AA}\) in XMEAM-V. The convergence criterion is relaxed to \(10^{-10} \text{ eV/\AA}\) for DP-HYB-V and \(10^{-6} \text{ eV/\AA}\) for GAP-V due to their higher computational cost and slower convergence.  The calculations of specific dislocation structures are described in Section~\ref{sec:calc_defect_method} below.

\subsection{EXtended modified embedded-atom method (XMEAM)}

The modified embedded-atom method (MEAM) was built upon the original embedded-atom method (EAM~\cite{daw_1984_prb}) by Baskes et al.~\cite{baskes_1987_prl,baskes_1989_prb} to describe bond-bending effects by including angular dependent terms in the formulation.
In MEAM, the total energy of a system of \(N\) atoms is
\begin{equation}\label{eq:meam_tot_e}
  E = \sum_{i}^{N}\left[ F_{i}\left( \bar{\rho}_{i} \right) + \dfrac{1}{2}\sum_{j\neq i}\phi_{ij}\left(r_{ij}\right)S_{ij}\right],
\end{equation}
 where \(F_{i},\phi_{ij},S_{ij}\) are the embedding function, pair interaction function and screening function. The embedding function usually takes the following form
  \begin{equation}
    F_{i}\left( \bar{\rho}_{i} \right) = A_{i}E_{i}^{0}\bar{\rho}_{i}\text{ln}\left( \bar{\rho}_{i} \right) + B_{i}\bar{\rho}_{i},
  \end{equation}
  where \(\bar{\rho}_{i}\) is the total background electron density at atomic site \(i\) due to all surrounding atoms and \(A_{i},E_{i}^{0},B_{i}\) are element-dependent parameters.  The contributions to \(\bar{\rho}_{i}\) in the original MEAM formulation include a spherically symmetric electron density term \({\bar{\rho}_i}^{(0)}\) and angular-dependent terms \(\bar{\rho}_i^{(k)}\). In the implementation of the LAMMPS~\cite{thompson_2022_cpc}), \(\bar{\rho}_{i}\) is expressed as
  \begin{equation}\label{eq:meam_e_dens_lammps}
    \bar{\rho}_i = \dfrac{\bar{\rho}_i^{(0)}}{\rho_i^0} G_i\left[ \sum_{k=1}^3 t_i^{(k)} \left( \dfrac{\bar{\rho}_i^{(k)}}{\bar{\rho}_i^{(0)}} \right)^2 \right]
  \end{equation}
  where \(G_i\) computes the electron density and has several forms~\cite{gullett_2003_sandia}, \(\rho_i^0\) is the composition-dependent electron density scaling, \(t_i^{(k)}\) are average weighting factors and \(k\) are truncated at 3.  The partial electron densities \(\bar{\rho}_{i}^{(k)}\) and average weighting factors \(t_{i}^{(k)}\) are parametrized by element-dependent \(\beta_i^{(k)}\) and \(t_{0,j}^{(k)}\) respectively (see Ref.~\cite{gullett_2003_sandia}).  Both are further multiplied by the screening function \(S_{ij}\)
  \begin{equation}
    S_{ij} = \prod_{k\neq i,j} S_{ikj} (C_\text{min}, C_\text{max}) f_\text{c}\left( \dfrac{r_\text{c}-r_{ij}}{\Delta r} \right)
  \end{equation}
  where \(C_\text{min}\) and \(C_\text{max}\) are the screening function parameters, \(f_\text{c}\) is the radial cutoff function, \(r_\text{c}\) is the cutoff distance and \(\Delta r\) the smoothing distance (see Refs.~\cite{gullett_2003_sandia,lee_2000_prb} for details).

  In the classical MEAM, the angular-dependent electron density in Eq.~\ref{eq:meam_e_dens_lammps} is truncated at \(k=3\) and only the first nearest-neighbor (1nn) interactions are explicitly treated.  The 1nn-MEAM exhibits difficulties in reproducing the ground state BCC/HCP structure and surface energy ordering of many elements.  Lee et al. thus modified the 1nn-MEAM to explicitly include the second nearest-neighbor (2nn) interactions~\cite{lee_2000_prb}, which successfully addressed these intrinsic issues in 1nn-MEAM.  The 1nn-MEAM/2nn-MEAM has been widely used to develop interatomic potentials for metals and semiconductors.  They have been shown to reproduce many properties of FCC~\cite{lee_2003_prb}, BCC~\cite{lee_2001_prb} and HCP~\cite{baskes_1994_msmse,kim_2006_prb} metals.  Despite its broad success, the 1nn/2nn-MEAM faces challenges in reproducing properties for multiple structures and their transition paths.  Multi-state MEAM (MS-MEAM~\cite{baskes_2007_prb}) was then introduced to address this shortcoming.  Specifically, MS-MEAM uses DFT-based multiple reference structures and transformation paths to determine all the functions/parameters in Eq.~\ref{eq:meam_tot_e}. The MS-MEAM was further expanded to include additional angular-dependent electron density functions (\(k=1,2,3,5\)) and was shown to well describe general properties of Ti in the HCP, BCC and liquid phases~\cite{gibson_2017_msmse}.

The MS-MEAM is particularly appealing for BCC TMs as it can include multiple reference structures (e.g., BCC, FCC and their transformation path).  Reproducing the relative energetics and their transition paths are critical for dislocation and twinning properties in BCC TMs.  We recently discovered that the \(\langle 111 \rangle/2\) screw dislocation core structure (ND vs D), its Peierls barrier \(\Delta E_\text{PB}\) and nucleation barrier \(\gamma_\text{us}\) are all related to the energy difference \(\Delta E_\text{FCC-BCC}\) bewteen the FCC and BCC structures~\cite{wang_2021}.  However, given the current deficiencies of DFT functionals in predicting some properties of V, it is possible that MS-MEAM will also inherit these deficiencies.  Here, we choose a new path by combining recent advances in MS-MEAM and the flexibility in the classical MEAM in choosing the target properties.  Specifically, the XMEAM in this work includes the angular electron density functions for \(k = 1,2,3,4,5\), which adds four parameters \(\beta^{4},\beta^{5},t^{4},t^{5}\) related to partial electron density functions.  In addition, the XMEAM uses independent screening parameters (\(C_{\text{min}}^{\text{rho}}, C_{\text{max}}^{\text{rho}}\))  in the electron density function and (\(C_{\text{min}}^{\text{pair}},C_{\text{max}}^{\text{pair}}\)) in the pair interaction function. The XMEAM thus has 6 more parameters in total, in addition to the 18 parameters in the classical MEAM. The XMEAM retains all the analytical functional forms and physical interpretations of the classical MEAM, and can be easily applied to a broad range of materials.  We have also developed the XMEAM based on the original implementation of MEAM in LAMMPS~\cite{thompson_2022_cpc}.  For the details of analytical expressions in XMEAM, we refer readers to earlier references~\cite{gullett_2003_sandia,lee_2000_prb,baskes_2007_prb,gibson_2017_msmse} and the source code available online~\cite{xmeam_2022_gitlab}. Here we focus on the fitting procedures of the XMEAM potential for V and its properties.

The datasets used for fitting potentials are critical for the accuracy and relevance of the resulting potentials.  For modelling the mechanical behaviour of V, we focus on the fundamental thermodynamic and mechanical properties of BCC V.  In particular, we use its BCC bulk properties (lattice parameter, cohesive energy, and elastic constants), vacancy and self-interstitial formation energies, equation of state (EOS), \(\{110\}\) surface decohesion energies and \(\gamma\)-lines of the \{110\}, \{112\}, and \{123\} planes.  In addition, we also include the cohesive energy and lattice parameter of the FCC phase in the fitting datasets, given its importance in governing dislocation properties~\cite{wang_2021}.  We use the experimental values of the BCC lattice parameter, cohesive energy and elastic constants, with the remainder from DFT as described in Section~\ref{sec:method_dft}.

\begin{table*}[!htbp]
	\caption{\label{tab:xmeam_v} Parameter fitting ranges and optimized values of the XMEAM-V developed in this work.}
	\centering
	\begin{threeparttable}
		\begin{tabular}{x{2.5cm} x{1.2cm} x{1.2cm} x{1.2cm} x{1.2cm}  x{1.2cm} x{1.2cm} x{1.2cm} x{1.2cm}}
			\toprule
      \textbf{Parameter} & \(\alpha\) & \(\beta^{0}\) & \(\beta^1\) & \(\beta^2\) & \(\beta^3\) & \(\beta^4\) & \(\beta^5\) & \(A\)\\
      \hline
      \text{Lower limit}     & 4.55   & 4.00   & 3.00    & 4.00   & 2.00    & 4.00    & 2.00   & 0.25   \\
      \text{Optimized value} & 4.6230 & 4.9371 & 3.8737  & 4.5265 & 3.6661  & 5.9816  & 2.6531 & 0.2965 \\
      \text{Upper limit}     & 4.65   & 6.00   & 5.00    & 6.00   & 4.00    & 6.00    & 4.00   & 0.35   \\
      \hline
      \textbf{Parameter}     & \(B\)  & \(t^{1}\) & \(t^2\) & \(t^3\)  & \(t^4\) & \(t^5\) & \(r_\text{c}\) & \(\Delta r\)\\
      \hline
      \text{Lower limit}     & -0.10  & -10.00    & 0.00    & -25.00   & -15.00  & -5.00   & 6.50      & 3.00   \\
      \text{Optimized value} & 0.0442 & -3.2282   & 5.1961  & -19.1000 & -8.6960 & 0.9093  & 7.6192    & 4.3217 \\
      \text{Upper limit}     & 0.10   & 0.00      & 10.00   & -15.00   & 0.00    & 5.00    & 8.00      & 5.00   \\
       \hline
      \textbf{Parameter}     & \(E_\text{c}\) & \(a_\text{lat}\) & \(d_{\text{attrac}}\) & \(d_{\text{repuls}}\) &  \(C_{\text{min}}^{\text{rho}}\) &
                              \(C_{\text{min}}^{\text{pair}}\) & \(C_{\text{max}}^{\text{rho}}\) & \(C_{\text{max}}^{\text{pair}}\) \\
      \hline
      \text{Lower limit}     & 5.28   & 2.98     & -0.10    & -0.10   & 0.00    & 0.30   & 2.60   & 2.60      \\
      \text{Optimized value} & 5.3000 & 0.2965   & 0.0119   & -0.0045 & 0.0454  & 0.4218 & 2.8616 & 2.9547    \\
      \text{Upper limit}     & 5.31   & 3.02     & 0.10     & 0.10    & 0.50    & 0.70   & 3.00   & 3.00      \\
      \bottomrule
		\end{tabular}
	\end{threeparttable}
\end{table*}

There are 24 adjustable parameters in an XMEAM potential for a pure element.  It is difficult to determine these parameters manually and yet achieve optimal properties.  We thus employ the particle swarm optimization (PSO~\cite{kennedy_1995_pieeeicnn,qin_2022_cms}) algorithm to optimize these parameters.  In the PSO algorithm, each candidate potential is a particle in a 24-dimensional space.  The potential parameters are thus the coordinates \(\mathbf{x}\) of the particle.  The properties of the candidate potential \(p_i\) are functions of this 24-parameter vector, i.e., \(p_i = f(\mathbf{x})\).

In practice, we assign an objective function
\begin{equation}\label{eq:pso_obj_func}
  f_\text{obj}(\mathbf{x}) = \sum_i^n w_i \left[ \dfrac{p_i - t_i}{t_i}  \right]^2
\end{equation}
where \(t_i\) and \(w_i\) are the target value and weight assigned to property \(i\).  The weights are chosen  based on relative importances of properties; high weightage is assigned to critical properties such as the surface energies of individual planes.  A large number of candidate potentials/particles are generated with their positions randomly chosen within a specified domain in the 24-dimensional space.  In the PSO process,  all the particles evolve based on their current positions, local optimal position and global optimal position (see Ref.~\cite{qin_2022_cms}).  The PSO process stops when the objective function \(f_\text{obj}\) falls below a threshold value or the number of iterations reaches a pre-defined limit.  We emphasize that a potential can rarely achieve perfect agreement with all target DFT/experimental values.  The PSO process thus yields an optimized potential within the fitting parameter space. Table~\ref{tab:xmeam_v} shows the parameter space/fitting range used for XMEAM-V.

\subsection{Deep potential hybridized (DP-HYB)}
The procedure to develop ML potentials is quite different from that of the classical potentials.  For DP-HYB-V, we use the general Deep Potential Generator (DP-GEN) scheme with the new hybrid descriptor~\cite{wang_2022_nf} and a ``specialization'' strategy~\cite{wen_2021_npjcm} to generate the training datasets.  The new hybrid descriptor includes two- and three-body functions modelled by embedding neural networks of sizes \((20, 40, 80)\) and \((4, 8, 16)\), respectively.  The fitting neural network size is \((240, 240, 240)\).  The cutoff radii for the two- and three-body embedding neural networks are 6 and 4 \AA, respectively.  In all the training stages, four models are trained on the same training datasets with the same neural network sizes but starting from different random seeds.

We first perform DFT calculations to determine the lattice parameters of the BCC, FCC and HCP structures of V (Section~\ref{sec:method_dft}).  Based on these lattice parameters, we construct three supercells of \(2\times2\times2\) BCC, FCC, and HCP structures containing 16, 32 and 16 atoms respectively.  These supercells are affinely scaled by \(\boldsymbol{s}\) from -4\% to 6\% with a step size of 2\%, resulting in 6 configurations for each phase. These scaled supercells are then perturbed 3 times by adding some random vectors \(\boldsymbol{\delta}=3\%\) to each of the supercell vectors \(\mathbf{c}_i\) and 0.01 {\AA} to atom positions \(\mathbf{R}_i\), which creates some distorted supercells.  \textit{Ab initio} MD (AIMD) simulations are then performed for 2 steps for each configuration. In AIMD, the NVT ensemble is employed with the temperature maintained at 100 K using the Nos\'e-Hoover thermostat.  At the end of AIMD steps, a total of 104 configurations from the converged ionic steps are prepared with atomistic information including the total energy, atom coordinates \(\mathbf{R}_i\), atomic forces \(\mathbf{f}_i\) and virial tensors. To enhance the description of the BCC structure near equilibrium, we create 20 more perturbations from each uniformly scaled BCC supercells and perform 5 AIMD steps, resulting in an additional 600 training datasets to represent the BCC V.

The above \textit{ab initio} configurations provide the initial training datasets to initialize the DP-GEN loop.  In the DP-GEN loop, 4 DP models are first generated randomly and trained using the initial datasets.  In each model, the learning rate starts at \(1\times10^{-3}\) and decays exponentially to \(5\times 10^{-8}\). In the DP-GEN iteration, the training step is \(4\times10^5\) and the pre-factors of the energy, atomic force, and virial tensor in the loss functions are \(p_{\text{e}}^{\text{start}} = 0.02\), \(p_{\text{e}}^{\text{limit}} = 2\), \(p_{\text{f}}^{\text{start}} = 1000\), \(p_{\text{e}}^{\text{limit}} = 1\), \(p_{\text{v}}^{\text{start}} = 0\), and \(p_{\text{v}}^{\text{limit}} = 0\), respectively.

In the exploration step of the DP-GEN loop, one DP model is selected to explore different bulk and surface structures using DP-based MD (DPMD) interfaced with the LAMMPS package.  We use fully periodic supercells of \(2\times 2 \times 2\) BCC, FCC and HCP structures and applied perturbations \(\boldsymbol{\delta}\).  The bulk configurations are explored using the NPT ensemble with fixed box shape.  The temperature and pressure are controlled using the Nos\'e-Hoover thermostat and barostat~\cite{martyna_1994_jcp,thompson_2022_cpc}.  For the bulk structures, DPMD explores 4 temperature range sets from 50 K to 1.9 $T_\text{m}$ ($T_\text{m}$ = 2183 K~\cite{rumble_2019_crc}):
\begin{equation}\label{eq:dp_temp_region}
\begin{aligned}
  &(a)\ 50 \ \text{K}, [0.1, 0.2, 0.3, 0.4] T_\text{m}; \\
  &(b)\ [0.5, 0.6, 0.7, 0.8, 0.9]T_\text{m}; \\
  &(c)\ [1.0, 1.1, 1.2, 1.3, 1.4]T_\text{m};  \\
  &(d)\ [1.5, 1.6, 1.7, 1.8, 1.9]T_\text{m}.
\end{aligned}
\end{equation}
In each temperature, DPMD are performed under 8 pressures [\(0.001, 0.01, 0.1, 1, 5, 10, 20, 50\)] kBar.

The exploration step provides a set of configurations based on the selected model.  The other 3 models are then used to calculate atomic force \(\mathbf{f}_i\) in these configurations, resulting in 4 sets of \(\mathbf{f}_i\).  The standard deviations \(\sigma(\mathbf{f}_i)\) is calculated and used as an indicator of the accuracy of the models.  If the maximum deviation of atomic forces \(\text{max}[\sigma(\mathbf{f}_i)]\) is within \([\sigma^{\text{low}}, \sigma^{\text{high}}]\), the configuration is considered a candidate and sent to DFT calculations. \(\sigma^{\text{low}}\) and \(\sigma^{\text{high}}\) are lower and upper bounds set as \([0.10,0.25]\) for region \((a)\), \([0.15,0.30]\) for regions \((b)\) and \((c)\), \([0.20,0.35]\) for region \((d)\) in Eq.~\ref{eq:dp_temp_region}.  All candidate configurations are computed with DFT and added to the initial datesets, forming a broad training datesets for the next DP-GEN loop.  The DP-GEN loop with bulk structures is iterated 32 times.

The DP-GEN loop, with the surface structures, follows the bulk structures exploration. For surface structures, the initial supercells are constructed for the \(\{100\}\), \(\{110\}\), and \(\{111\}\) surfaces in the BCC and FCC structures, and the \(\{0001\}\) and \(\{10\bar{1}0\}\) surfaces in the HCP structure. The surface structures are then uniformly scaled by \(\boldsymbol{s}\) and perturbed by \(\boldsymbol{\delta}\).  The surface configurations are explored within the NVT ensemble from 50 K to 0.9 $T_\text{m}$ (temperature regions \((a)\) and \((b)\) in Eq.~\ref{eq:dp_temp_region}).  For exploring the surface structures, \(\sigma^{\text{low}}\) and \(\sigma^{\text{high}}\) are \(0.20\) and \(0.35\) in the entire temperature range.  The DP-GEN loop with surface structure is iterated 8 times.

In addition to the bulk and surface structure datasets, we also include configurations with point defects in the BCC structure, which are important for diffusion, vacancy and interstitial clustering.  We include 6 types of self-interstitials (Fig.~\ref{fig:self_interstitial}).  Specifically, we compute the \(\langle 111 \rangle\), \(\langle 110 \rangle\) and \(\langle 100 \rangle\) dumbbells, \(\langle 111 \rangle\) crowdion, tetrahedral, and octahedral interstitial structures in a \(3\times 3\times3\) supercell (55 atoms) using DFT.  For each self-interstitial, a set of configurations are obtained via ionic structure optimization.  These configurations are used as initial configurations for DPMD exploration using one selected DP model, in the temperature range of 50 to 600 K and the NVT ensemble.  The other 3 DP models are then used to calculate the atomic forces in these DPMD configurations. The configurations with \(\text{max}[\sigma(\mathbf{f}_i)]\) within \([0.2,0.35]\) are selected as candidates and sent for DFT calculations.  These point-defect related \textit{ab initio} configurations are added to the training datasets. The DP-GEN loop with self-interstitial structures is iterated 8 times.

Finally, we include configurations with atoms at small separations in the BCC and FCC structures.  We create \(2\times 2\times2\) BCC and FCC cubic supercells and uniformly scale the supercell vectors from \(0.86\) to \(0.94\) with a step size of \(0.02\).  These supercells are further distorted by adding \(\boldsymbol{\delta}\) to its supercell vectors and atomic positions.  The distorted supercells are used as initial configurations for DPMD with one selected DP model.  DPMD is then performed with the NVT ensemble at 50 to 600 K for exploration at high atomic density/small atom separations.  Similar to the earlier DP-GEN loop, the forces are computed in the DPMD configurations with the other 3 DP models.  The DPMD configurations with \(\text{max}[\sigma(\mathbf{f}_i)]\) within \([0.2,0.35]\) are selected as candidates and sent for DFT calculations.  The resulting configuration from DFT are added to the training datesets. The DP-GEN loop with high density atomic structures is iterated 4 times.

The above DP-GEN loops explore configurations in the bulks, at the surfaces, with point defects and at high densities.  They provide a broad set of atomic configurations/training datasets relevant to mechanical properties of V.  Table~\ref{tab:dp_training_data} summarizes the training datasets generated in the DP-GEN Loop.  We refer readers to Refs.~\cite{han_2018_ccp,wen_2021_npjcm} for more details of the DP-GEN scheme.

\begin{table*}[!htbp]
	\caption{\label{tab:dp_training_data} Summary of the training datasets for DP-HYB-V.}
	\centering
	\begin{threeparttable}
		\begin{tabular}{x{6cm} x{4cm} x{1.5cm}}
			\toprule
			\textbf{Dataset type} & \textbf{Number of datasets} & \textbf{Weightage} \\
			\hline
			Initialization datasets around equilibrium & 704 & 1 \\
			DP-GEN bulk & 3393 & 1 \\
			DP-GEN surface & 991 & 1 \\
			DP-GEN interstitial & 705 & 1 \\
			DP-GEN high-density BCC & 1052 & 1 \\
			DP-GEN high-density FCC & 1846 & 1 \\
			$\gamma$-line datasets from specialization & 63 & 100 \\
			Cohesive energy datasets from specialization & 10 & 100 \\
			Vacancy datasets from specialization & 22 & 10 \\
			Total & 8786 &  \\
			\bottomrule
		\end{tabular}
	\end{threeparttable}
\end{table*}

Based on the above DP-GEN loop, the 4 DP models can reproduce many properties of BCC V.  However, their generalized stacking fault energies (\(\gamma\)-lines) are not sufficiently accurate with respect to DFT results.  Special training datasets are thus generated with configurations describing shear displacement along the \(\langle 111 \rangle\) direction on the \(\{110\}, \{112\}, \text{and} \{123\}\) planes in the BCC structure.  In addition, ``special'' training sets are generated on the EOS curve of BCC V (Fig.~\ref{fig:coh_vs_a}) for a wide range of lattice parameters \(a/a_{0} = 0.75, 1.2, 1.3,..., 2.0\) (\(a_{0}\) is the equilibrium lattice parameter) and for a monovacancy configuration in a supercell of \(3 \times 3 \times 3\) BCC V.  For each configuration, DFT is employed to compute the total energy and atomic forces, which form the special datasets and are added to the earlier training datasets.  The final training datasets include those from the initial, DP-GEN loop and special steps (Table~\ref{tab:dp_training_data}).  Four new DP-HYB models are trained with \(8\times 10^6\) steps using all the training sets in Table~\ref{tab:dp_training_data} (except the high-density FCC structures and cohesive energy datasets) with a focus on the properties of BCC structures near equilibrium.  Subsequently, the best performing DP-HYB model is further trained with all the training sets in Table~\ref{tab:dp_training_data} for \(4\times 10^6\) steps with the initial and final learning rates at \(1\times 10^{-4}\) and \(5\times 10^{-8}\), respectively.  The pre-factors in both training processes are \(p_{\text{e}}^{\text{start}} = 10\), \(p_{\text{e}}^{\text{limit}} = 10\), \(p_{\text{f}}^{\text{start}} = 1\), \(p_{\text{f}}^{\text{limit}} = 1\), \(p_{\text{v}}^{\text{start}} = 10\), and \(p_{\text{v}}^{\text{limit}} = 10\), respectively.  Table~\ref{tab:dp_training_data} summarizes the individual weights assigned to the respective datasets for the final training for DP-HYB-V.

Finally, the energy of DP-HYB-V is adjusted with respect to the energy of an isolated V atom in vacuum, in order to reproduce its cohesive energy of BCC V (\(-5.31\) eV/atom~\cite{kittel_2019_issp}) measured in the experiment.  This adjustment was also applied in DP-Ti developed earlier~\cite{wen_2021_npjcm}.


\subsection{Calculation of lattice properties and defects}
\label{sec:calc_defect_method}

A wide range of lattice and defect properties are calculated using the developed XMEAM-V and DP-HYB-V.  The point defects are calculated using the same setup as that in DFT~\ref{sec:method_dft}.  We describe the details of the calculations which are sensitive to the simulation conditions below.

\begin{table*}[!htbp]
  \caption{\label{tab:v_dis_sc} Supercell orientations and sizes used to calculate Peierls stresses of different dislocations.}   \centering
  \begin{threeparttable}
    \begin{tabular}{x{4cm} x{4cm} x{4cm} x{4cm}}
      \toprule
      \textbf{Slip system}  & \textbf{Supercell orientation} & \textbf{Size} & \textbf{Number of atoms} \\
            \(\mathbf{b}, \mathbf{n}\)                     & \((\mathbf{c}_1,\mathbf{c}_2,\mathbf{c}_3)\)& \((\left \lvert  \mathbf{c}_1 \right  \rvert, \left \lvert  \mathbf{c}_2 \right \rvert, \left \lvert  \mathbf{c}_3 \right \rvert)\) &  \\
      \hline
      \(1/2\langle 111 \rangle \) screw & \(([111],[11\bar{2}],[1\bar{1}0])\) & (\(\sqrt{3}a_{0},54\sqrt{6}a_{0},47\sqrt{2}a_{0}\)) & 30456\\
      \(1/2\langle 111 \rangle \{110\} 70.5^\circ\) mixed & \(([111],[11\bar{2}],[1\bar{1}0])\) & \((\sqrt{3}a_{0},54\sqrt{6}a_{0},47\sqrt{2}a_{0})\)  & 30456 \\
      \(1/2\langle 111 \rangle \{110\}\) edge & \(([111],[11\bar{2}],[1\bar{1}0])\) & \((154\sqrt{3}/2a_{0},\sqrt{6}a_{0},47\sqrt{2}a_{0})\) & 43428 \\
      \(\langle 100 \rangle \{110\}\) edge  & \(([100],[011],[01\bar{1}])\) & \((133a_{0},\sqrt{2}a_{0},46\sqrt{2}a_{0})\) & 24472\\
      \bottomrule
    \end{tabular}
  \end{threeparttable}
\end{table*}

We use a periodic array of dislocations (PAD~\cite{bulatov_2006_csd}) configuration to investigate all dislocation core properties in this work.  In the PAD configuration, the slip plane is placed in the \(x-y\) plane and normal to the \(z\)-direction.  Periodic boundary conditions are imposed in the \(x\) and \(y\) directions, while the top and bottom \(z\) surfaces are treated as traction-controlled/free surfaces.  For the pure edge and screw dislocations, the Burgers vector \(\mathbf{b}\) is aligned in the \(x\) direction. For mixed dislocations, the screw component of \(\mathbf{b}\) (i.e., \(\left \lvert \mathbf{b} \cdot \boldsymbol{\xi} \right \rvert \boldsymbol{\xi}\), where \(\boldsymbol{\xi}\) is the dislocation line direction) is always aligned in the \(x\) direction.  Shear stresses are applied by adding forces to atoms at the top and bottom layers within 12 \(\text{\AA}\) \((\sim 2\times r_\text{c}\), where \(r_\text{c}\) is the cutoff distance of the interatomic potential) from the surfaces in the \(\pm z\)-directions.

We first construct the respective supercells with the appropriate crystal orientations and dimensions, as shown in Table~\ref{tab:v_dis_sc}.  Dislocations are then introduced at the center of the supercell by applying the displacement field of the corresponding Volterra dislocation using the atomsk package~\cite{hirel_2015_cpc}.  For dislocations with a nonzero screw component, a homogeneous shear strain of \(\epsilon_{yx} = \mathbf{b} \cdot \boldsymbol{\xi} /2\) is applied to correct the plastic shear strain created by the screw component.  The constructed cores are then optimized using the conjugate gradient algorithm with a force convergence criterion of \(10^{-4} \text{ eV/\AA}\).

In the calculations of the Peierls stress, we use a load-optimize sequence to estimate the critical stress to drive the dislocation at 0 K. In each case, some trial runs are firstly carried out to estimate \(\tau_\text{P}\), where \(\tau_\text{P}\) is the final stress when the dislocation starts to glide continuously.  In the actual measurement of the Peierls stresses, we first apply a shear stress \(\tau_{\text{step}}\) of about \(5\)\%\(\tau_\text{P}\).  Structure optimization is then carried out with the applied stresses/forces fixed.  Upon reaching convergence/equilibrium, the shear stress is increased by \(\tau_\text{step}\), followed by structure optimization.  The load-optimize sequence is repeated until the dislocation starts to glide continuously.  When the system nearly approaches the final \(\tau_{\text{P}}\), \(\tau_{\text{step}}\) is reduced to less than \(1\)\%\(\tau_{\text{P}}\).  The Peierls energy profile for the \(\langle 111 \rangle/2\) screw dislocation is calculated using the nudged elastic band (NEB~\cite{jonsson_1998_cqdcps}) method with the force convergence criterion set at \(10^{-6} \text{ eV/\AA}\).

For the \(\langle 111 \rangle/2\) screw dislocation, the measured \(\tau_\text{P}\) is weakly influenced by the simulation supercell sizes.  For example, \(\tau_\text{P}\) of a screw dislocation of length \(2\left \lvert  \mathbf{b} \right \rvert\) in XMEAM-V is measured as 1148 MPa, 1210 MPa and 1217 MPa in supercells of sizes (\(200\text{ \AA}\), \(80\text{ \AA}\)), (\(300\text{ \AA}\), \(140\text{ \AA}\)) and (\(400\text{ \AA}\), \(200\text{ \AA}\)).  The core structures and Peierls barriers are almost identical in the different supercell sizes.  Consequently, we use the smaller supercell of (\(200\text{ \AA}\), \(80\text{ \AA}\)) to calculate the Peierls barrier and related screw core energetics, and (\(400\text{ \AA}\), \(200\text{ \AA}\)) to calculate the Peierls stress for the screw dislocation and other types of dislocations. The supercell sizes are summarized in Table~\ref{tab:v_dis_sc}.

The temperatures-dependent lattice parameters \(a({T})\) or elastic constants \(C_{ij}(T)\) are calculated by time-averaging the supercell sizes using an NPT ensemble or time-averaging the measured stresses using the NVT ensemble equilibrated with a Nos\'e-Hoover thermostat in LAMMPS. Specifically, fully periodic supercells are used for the calculations. The supercell sizes are \(24\times 24 \times 24\) for XMEAM-V, while a smaller supercell of \(12\times 12 \times 12\) are used for GAP-V and DP-HYB-V due to their higher computational cost.  For the calculation of \(a({T})\), the supercell is first equilibrated for 40000 fs (40000 time steps) under stress-free conditions and at the target temperature with an isothermal-isobaric NPT ensemble.  The supercell size is then measured and averaged for 4000 steps.  The final supercell size is averaged for 10 measurements.

For the calculation of \(C_{ij}({T})\), the supercell is first equilibrated at target temperatures using an NPT ensemble for 16000 time steps. A \(\pm1\%\) strain in one of the strain components (\(\epsilon_{11},\epsilon_{22},\epsilon_{33},\epsilon_{12},\epsilon_{13},\epsilon_{23}\)) is applied and the configuration is then equilibrated for another 16000 steps using the canonical NVT ensemble.  After the equilibration, the resulting stresses are measured and averaged for every 4000 times steps.  The final stresses are taken as the average of 4 measurements. For DP-HYB-V, \(C_{44}\) based on the smaller \(12\times 12 \times 12\) supercell exhibits large fluctuations at high temperatures.  We therefore perform additional calculations using a \(24\times 24 \times 24\) supercell for temperatures above 1100 K.  The elastic constants are calculated by dividing the measured stresses with the applied strains.

We use PHONOPY~\cite{togo_2015_sm} and phonoLAMMPS~\cite{phonolammps_2022_github} to calculate the phonon spectra of the BCC structure at 0 K. The supercell size is \(8[100] \times 8[010] \times 8[001]\), containing 1024 atoms in total. Specifically, phonoLAMMPS is first used to compute the \(3N \times 3N\) force matrix and PHONOPY is used to calculate the phonon spectra based on the force matrix.

The melting temperatures is determined by the solid-liquid two-phase co-existence method. We employ a fully periodic supercell of \(92\text{\AA} \times46\text{\AA}\times46\text{\AA}\) and examine the volume fraction of the liquid and BCC phases in the supercell as a function of temperature in the NPT ensemble.  The melting point is determined to be the temperature above which the BCC phase grows and below which the BCC phase shrinks.  Separately, we also estimate the melting temperature in the calculations of elastic constants \(C_{ij}(T)\) when the shear modulus \(C_{44}\) drops to zero, as shown below.

\section{Results}
\label{sec:results}

Table~\ref{tab:xmeam_v} shows the final optimized parameters of XMEAM-V. The potential files in the LAMMPS format are also available in the Supplementary Materials. DP-HYB-V and all the training datasets are available at dplib~\cite{dphybv_2022_dplib}.  Both XMEAM-V and DP-HYB-V are compatible with LAMMPS and can be employed immediately.

In this section, we provide a comprehensive study of the classical (XMEAM-V) and ML (DP-HYB-V) potentials on their thermodynamic and mechanical properties.  We perform extensive benchmarks using GAP-V, DFT and experimental results of V available in the literature.

\subsection{Bulk properties}

\begin{table*}[!htbp]
\caption{\label{tab:basic_prop} Properties of V from DFT, experiments, XMEAM-V, DP-HYB-V and GAP-V. The properties include lattice parameter \(a\) (\AA), cohesive energy \(E_{\text{c}}\) (eV), elastic constants (GPa) in BCC and FCC phases, and surface energies \(\sigma\) (J/m\(^{2}\)) of low index planes in BCC phase. DFT calculations are performed in this work. }
\centering
\begin{threeparttable}
\begin{tabular}{x{2cm} x{1cm} x{2cm} x{2.5cm} x{2cm} x{2cm} x{2cm} x{2cm}}
\toprule
  \textbf{Structure} & \multicolumn{2}{c}{\textbf{Property}}  & \textbf{Experiment} & \textbf{DFT} & \textbf{XMEAM-V} & \textbf{DP-HYB-V} & \textbf{GAP-V}~\cite{byggmastar_2020_prm} \\
\hline
  \multirow{9}{*}{BCC} & \(a_0\)   &        & \textbf{3.03} (300 K)~\cite{kittel_2019_issp} &  3.00      & 3.03 & 3.00 & 3.00     \\
            & \(E_\text{c}\)   &      & \textbf{-5.31}~\cite{kittel_2019_issp} &  -5.38      &  -5.323         &    \(-\)5.308 &   -5.384      \\
            &                         & \(C_{11}\)  & \textbf{232.4} (0 K~\tnote{a} )~\cite{simmons_1971_mit} &  268.6 (15.6\%) &  261.4 (12.5\%)          &    271.9 (17.0\%)  &   271.0 (16.6\%)      \\
            & \(C_{ij}\)              & \(C_{12}\)  & \textbf{119.4}~\cite{simmons_1971_mit} &  140.0 (17.3\%) &  104.2 (12.7\%)          &    141.6 (18.6\%)  &   145.0 (21.4\%)      \\
            &                         & \(C_{44}\)  & \textbf{46.0}~\cite{simmons_1971_mit} &  23.6 (48.7\%)   &  41.3 (10.2\%)          &    25.3 (45.0\%) &   23.7 (48.5\%)       \\
                     &                         & \{100\}      & \multirow{4}{*}{\textbf{2.62}\tnote{b} ~\cite{tyson_1977_ss},\textbf{2.55}\tnote{b}~\cite{boer_1989_cmtma}} &  2.39             &  2.66              &    2.60   &   2.38     \\
            &                         & \{110\}      &  &  2.41             &  2.36             &    2.35   &   2.40      \\
            & \(\sigma\)              & \{112\}      &  &  2.71             &  2.72            &    2.63   &   2.69       \\
            &                         & \{123\}      &  &  2.64             &  2.66              &    2.60    &   2.64    \\
      \hline
  \multirow{6}{*}{FCC} & \(a\)                        &   &      &       3.82        &  3.85              &   3.82  &  3.82        \\
            & \(E_{c}\)               &          &         &     -5.14       &  -5.145            &   -5.064   &  -5.142     \\
            & \multicolumn{2}{l}{\(\Delta E_{\textbf{FCC-BCC}}\)(\(\chi\))} &   &  0.243       &  0.178 (0.74) &  0.244 (1.00) & 0.242 (1.00) \\
            &                         & \(C_{11}\)   & &  4.7                &  97.3                &   8.1   & 16.9      \\
            & \(C_{ij}\)              & \(C_{12}\)  & &  262.4               &  180.8              &   276.1  & 265.9       \\
            &                         & \(C_{44}\)   & &  5.4                &  44.5              &   1.2   & 9.2         \\
      \bottomrule
\end{tabular}

\begin{tablenotes}
\item[a] Experimental elastic constants at 0 K are extrapolated from a series of measurements at low temperatures.
\item[b] Experimental surface energies are obtained from surface tension measurements and do not represent the property of a specific surface.
\end{tablenotes}
  \end{threeparttable}
\end{table*}

Table~\ref{tab:basic_prop} shows the basic properties of BCC and FCC V calculated by DFT, XMEAM-V, DP-HYB-V and GAP-V, as well as the experimental data of BCC V.  For the BCC structure, all computational models accurately reproduce the lattice parameter, cohesive energy, surface energies and elastic constants \(C_{11}\) and \(C_{12}\).  Nevertheless, DFT predicts \(C_{44}\) as 23.6 GPa, about 50\% of the experimental value of 46.0 GPa and consistent with most ealier DFT calculations for V~\cite{byggmastar_2020_prm,li_2020_npjcm}. ML DP-HYB-V and GAP-V faithfully reproduce this value from DFT.  We attempted to manually correct the \(C_{44}\) value by scaling the virial tensor relevant to \(C_{44}\) in the DP-HYB training datasets. However, this strategy leads to degradation of other properties like the dislocation core structures.  Manually adjusting \(C_{44}\) perhaps generates some inconsistencies among the training datasets, and this turns out to be not as straightforward as anticipated. On the contrary, XMEAM-V reproduces all the elastic constants within 15\% from their respective experimental values.

For the FCC phase, V is not stable at 0 K as predicted by DFT. Its \(C_{11}\) and \(C_{44}\) are nearly 0 and \(C_{11} < C_{12}\), which violates the Born stability criterion for cubic structures~\cite{mouhat_2014_prb}. No FCC phase appears in the low pressure regions of the equilibrium phase diagram of V, further suggesting the instability of FCC-V. The two ML potentials also show excellent reproducibility of the FCC elastic constants from DFT, while the XMEAM-V shows appreciable discrepancies.  DFT also predicts that the FCC phase has a higher cohesive energy relative to the BCC phase, \(\Delta E_{\text{FCC-BCC}} = 0.243 \text{ eV/atom}\). The two ML potentials accurately reproduce this quantity, while XMEAM-V has \(\Delta E_{\text{FCC-BCC}} = 0.178 \text{ eV/atom}\) or \(\chi = \Delta E^{\text{XMEAM-V}}_{\text{FCC-BCC}} /\Delta E^{\text{DFT}}_{\text{FCC-BCC}} = 0.74\). Further increase of \(\Delta E_{\text{FCC-BCC}}\) in XMEAM-V is feasible, but at a cost of reduced accuracy in other properties such as the \(\gamma\)-lines. The current XMEAM-V is thus selected with an overall balance on all properties considered.  Comparing the two approaches, ML potentials demonstrate excellent capability in reproducing multi-phase/state properties while the classical approach exhibits some limitations here.

\subsection{Equation of state}

\begin{figure}[!htbp]
  \centering
  \includegraphics[width=0.495\textwidth]{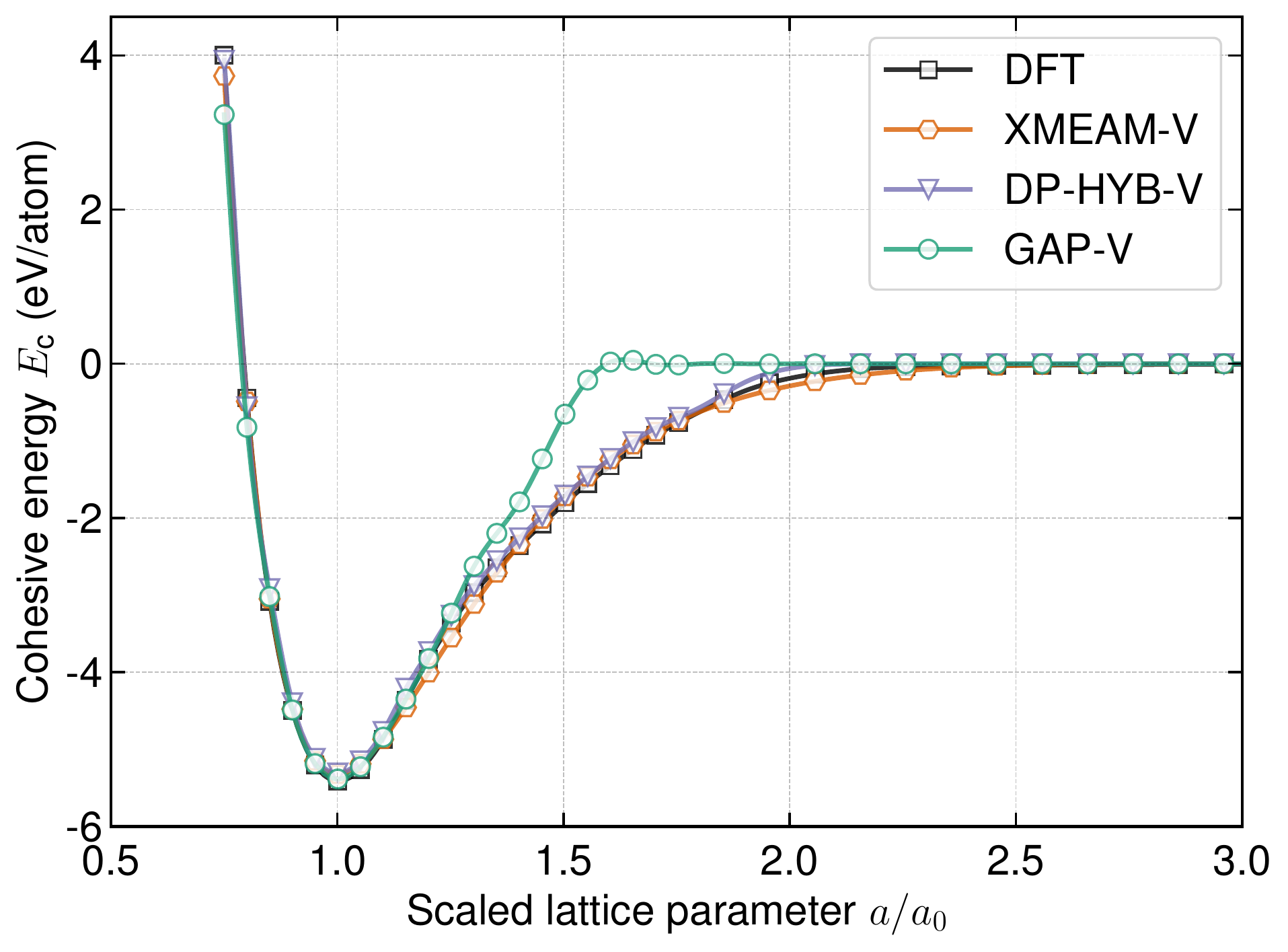}
  \caption{\label{fig:coh_vs_a} The cohesive energy as a function of lattice parameter of the BCC phase of V predicted by DFT and three interatomic potentials (XMEAM-V, DP-HYB-V and GAP-V~\cite{byggmastar_2020_prm}).}
\end{figure}

Figure~\ref{fig:coh_vs_a} shows the BCC cohesive energy as the function of the lattice parameter (i.e., equation of state, EOS) predicted by DFT and the interatomic potentials.  All three potentials accurately reproduce the DFT values near the equilibrium lattice parameter in the range \((0.8a_{0},1.2a_{0})\).  XMEAM-V and DP-HYB-V have smooth energy variations in the entire range \((0.7a_{0},3.0a_{0})\), while GAP-V shows a rapid energy increase near and beyond \(1.4a_{0}\).  This rapid change in cohesive energy is perhaps due to its relatively short cutoff distance of \(4.7\text{ \AA}\) employed in the potential, as seen in many other MEAM-type interatomic potentials~\cite{wu_2015_msmse}.  While the physical implication of the rapid energy variation at large atom separations is not completely clear, it does affect other fundamental properties such as the traction-decohesion behaviour of some low-index planes, as shown below.

\subsection{Surface energy and decohesion}
All computational models accurately capture the surface energies of BCC V when compared to values measured by the surface tension method~\cite{tyson_1977_ss,boer_1989_cmtma}.  DFT predicts that the \(\{100\}\) plane has the lowest surface energy, followed by the \(\{110\}\), \(\{123\}\) and \(\{112\}\) planes.  GAP-V reproduces all surface energies and their ordering from DFT, while XMEAM-V and DP-HYB-V show that the \(\{110\}\) plane has the lowest energy, followed by the \(\{100\}\), \(\{123\}\) and \(\{112\}\) planes. We note that for all other BCC TMs, the close-packed \(\{110\}\) plane has the lowest energy and V is the only exception~\cite{byggmastar_2020_prm}.  Nevertheless, the energy differences among the different planes are rather small and the largest discrepancies are 11.3\% and 8.8\% on the \(\{110\}\) plane in XMEAM-V and DP-HYB-V.  All models predicts the \(\{100\}\) and \(\{110\}\) planes as the primary cleavage planes in BCC V~\cite{tyson_1973_am}.

\begin{figure*}[!htbp]
  \centering
  \includegraphics[width=0.9\textwidth]{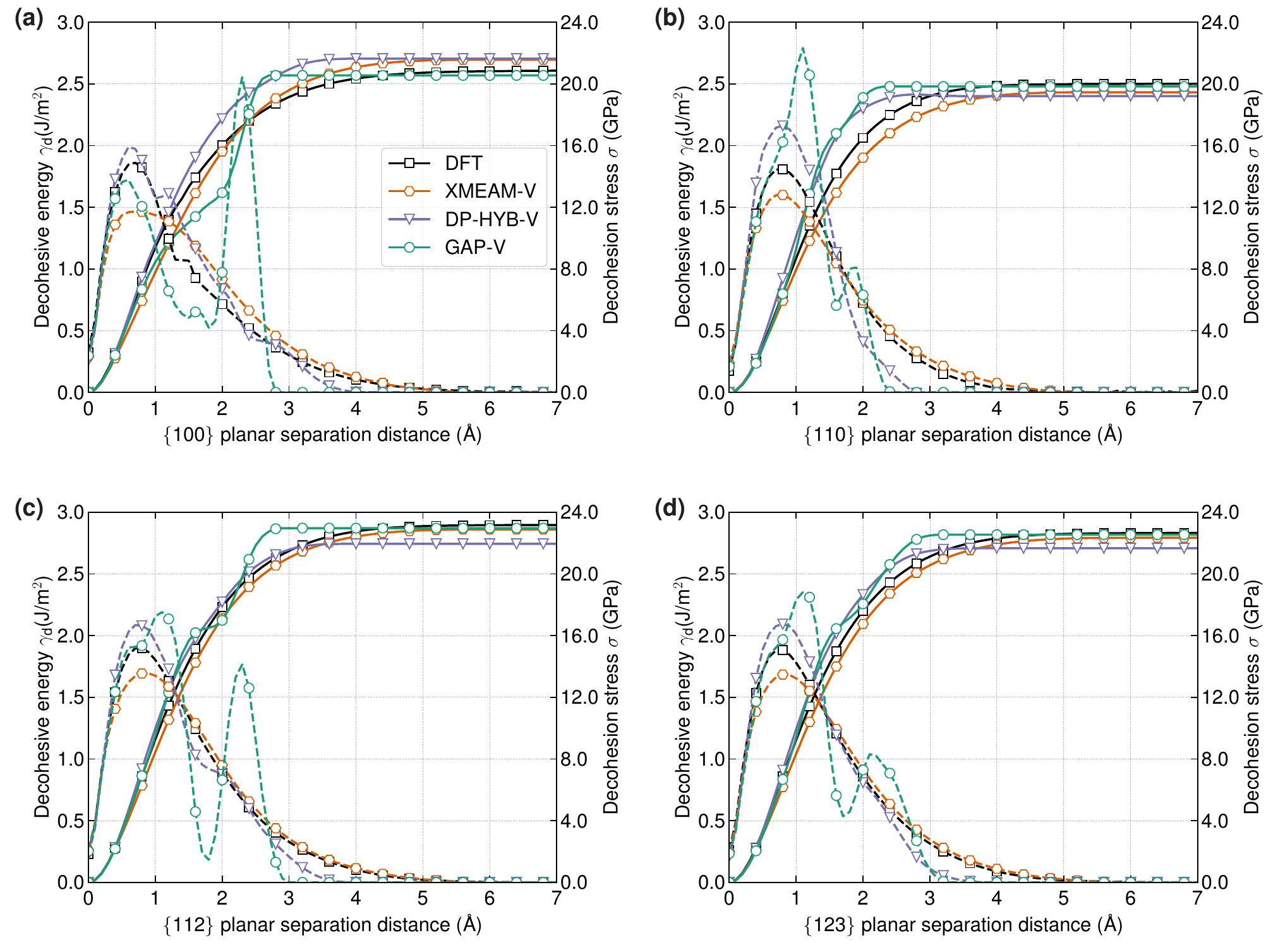}
  \caption{\label{fig:deco_vs_a} Surface decohesion energy curves (solid) and their gradients/stresses (dashed) of the \(\{100\}, \{110\}, \{112\}\), and \(\{123\}\) planes predicted by DFT and three potentials. Note that the decohesion energies are divided by 2 and thus the energy at large distance is the unrelaxed surface energy. }
\end{figure*}

BCC V tends to be brittle and exhibits cleavage fracture at low temperatures~\cite{joseph_2007_jnm}. The cleavage process is governed by the surface traction-separation relations.  We compute the surface decohesion energy by rigidly separating two blocks of materials across a specified crystallographic plane.  Figure~\ref{fig:deco_vs_a} shows the results obtained from DFT and the three interatomic potentials.  For XMEAM-V, the DFT data of the \(\{110\}\) plane is included in the fitting datasets, while the DP-HYB-V and GAP-V do not explicitly include these data.  All 4 low-index planes are included for comparisons.  Both XMEAM-V and DP-HYB-V capture the decohesion energy variations, gradients and peak values from DFT, while GAP-V exhibits undulating decohesion stresses at large planar separations.  These undulations occur at planar separation approaching \(2 \text{ \AA}\) where atoms move outside their interaction distance in GAP-V, as that in the EOS curves in Fig.~\ref{fig:coh_vs_a}.

\subsection{Generalized stacking fault energy}

The generalized stacking fault energy (\(\gamma\)-surface) describes the periodic energy variations during shear displacement between two crystallographic planes. The \(\gamma\)-surface has fundamental importance in governing dislocation nucleation, dissociation, core structure, energy and glide behaviour.  The minimum energy path bewteen two absolute minima on the \(\gamma\)-surface is the \(\gamma\)-line and is the fundamental slip step during plastic deformation.  Along the \(\gamma\)-line, the maxima is the unstable stacking fault energy \(\gamma_{\text{usf}}\) and dictates the dislocation nucleation barrier at stress concentrations such as crack-tips, while the metastable point \(\gamma_{\text{sf}}\) determines the dislocation core dissociation.

\begin{figure*}[!htbp]
  \centering
  \includegraphics[width=0.95\textwidth]{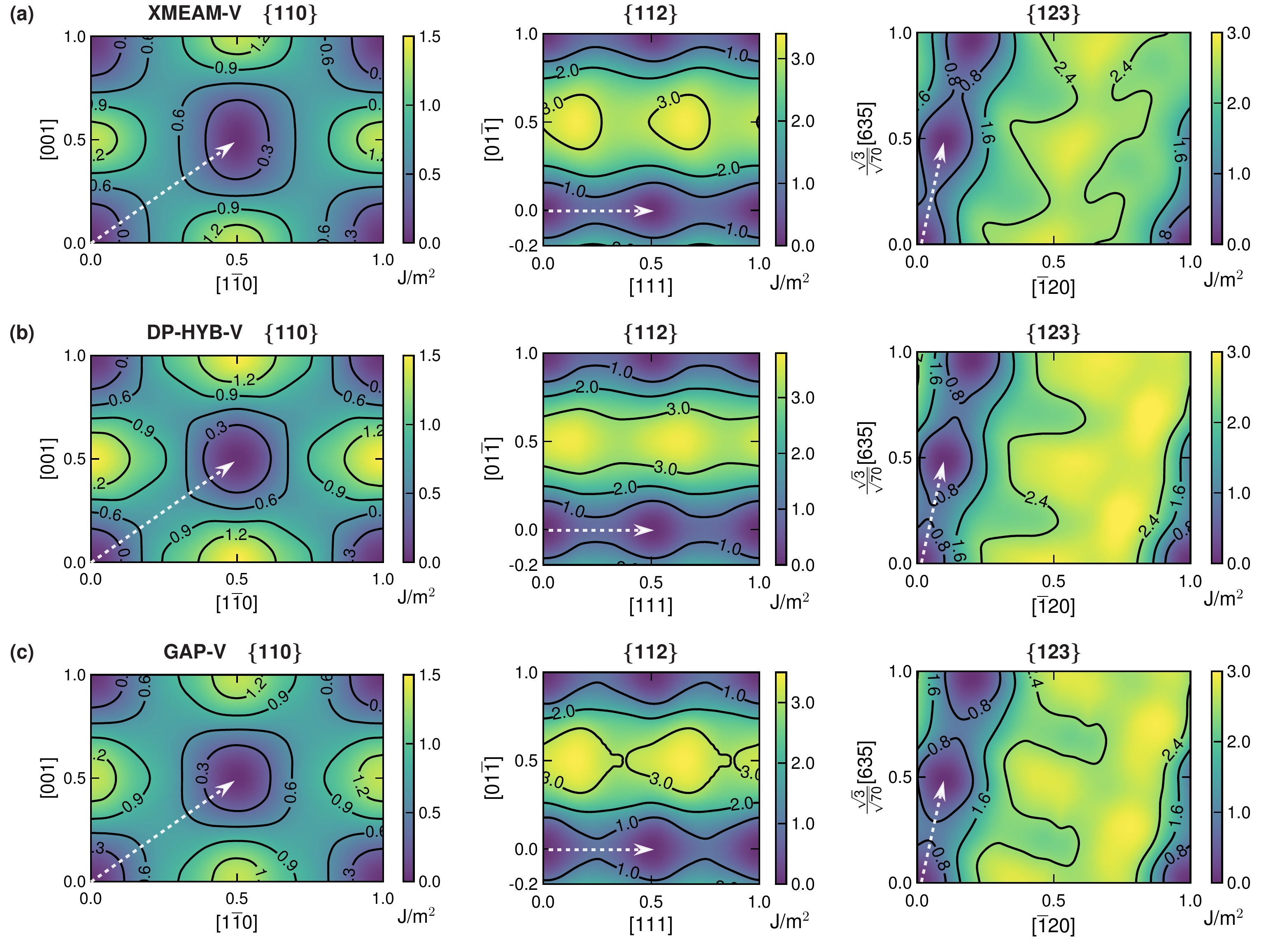}
  \caption{\label{fig:pot_gamma_surface} Generalized stacking fault energy surfaces (\(\gamma\)-surfaces) of the \(\{110\}\), \(\{112\}\), and \(\{123\}\) planes calculated by XMEAM-V, DP-HYB-V and GAP-V. The white dashed arrows denote the shortest lattice translation vector \(1/2\langle 111 \rangle\) on the respective planes.}
\end{figure*}

Figure~\ref{fig:pot_gamma_surface} shows the \(\gamma\)-surfaces of the \(\{110\}\), \(\{112\}\) and \(\{123\}\) planes in the BCC structure predicted by XMEAM-V, DP-HYB-V and GAP-V. These three planes are the common active slip planes in BCC metals. In all the cases, the minimum energy path is along the \(\langle 111 \rangle\) direction.  No metastable point exists in any of the cases, which is consistent with the \(\gamma\)-surfaces of other BCC metals computed by DFT~\cite{wang_2021,romaner_2010_prl}.  All \(\langle 111 \rangle/2\) dislocations (screw, edge and mixed) are thus expected to have non-dissociated core structures.  Despite the completely different potential energy functions and fitting methods, all three models exhibit similar \(\gamma\)-surface profiles with some differences near the peak energies. The \(\gamma\)-surfaces are perhaps strongly dictated by crystal geometry which is easily represented in all models.

\begin{figure*}[!htbp]
  \centering
  \includegraphics[width=0.9\textwidth]{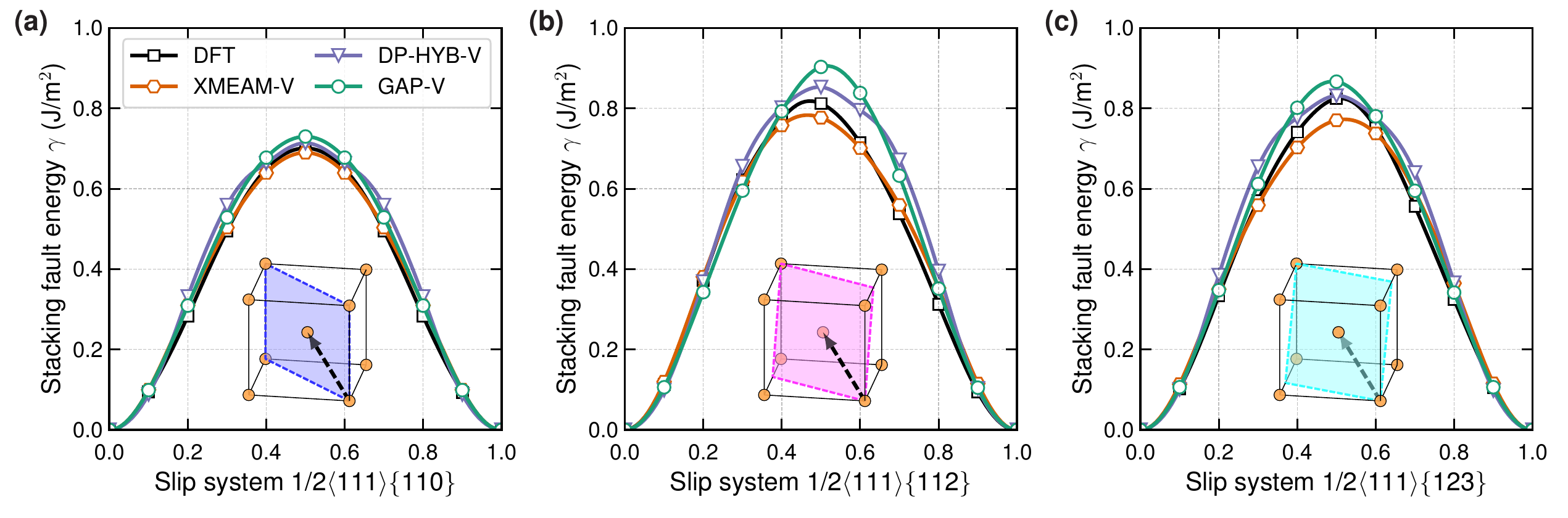}
  \caption{\label{fig:gamma_line} Generalized stacking fault energy lines (\(\gamma\)-lines) along the \(\langle 111 \rangle\) direction on the \(\{110\}, \{112\}\), and \(\{123\}\) planes predicted by DFT, XMEAM-V, DP-HYB-V and GAP-V. }
\end{figure*}

Figure~\ref{fig:gamma_line} shows the \(\gamma\)-lines along the \(\langle 111 \rangle/2\) direction on the three planes calculated by DFT and the respective interatomic potentials.  These \(\gamma\)-lines are included in the fitting datasets of XMEAM-V, DP-HYB-V and GAP-V.  The \(\{110\}\) plane is the close-packed plane and has the largest interplanar separation, followed by the \(\{112\}\) and \(\{123\}\) planes.  The DFT-based unstable stacking fault energies exhibit a similar trend, i.e., \(\gamma_\text{us}^{\{110\}} < \gamma_\text{us}^{\{112\}} \approx \gamma_\text{us}^{\{123\}}\).  All interatomic potentials accurately reproduce the \(\gamma\)-line profiles and their peaks; the largest discrepancy is 11\% of GAP-V on the \(\{112\}\) plane.  Overall, XMEAM-V possesses accurate elastic constants, surface-decohesion lines and \(\gamma\)-lines. It is thus promising to be used to study general plastic and fracture behaviour of BCC V.

\subsection{Point defects}
\begin{table*}[!htbp]
  \caption{\label{tab:v_point_defect} The formation and migration energies (eV) of the monovacancy and formation energies of self-interstitials of BCC V based on experiments, DFT, XMEAM-V, DP-HYB-V and GAP-V. In MD simulations with the relaxation on atomic positions and supercell vectors, some self-interstitial configurations are unstable and relax to lower energy states.  The numbers in the parenthesis are energies obtained under fixed supercell constraint.}
  \centering
  \begin{threeparttable}
    \begin{tabular}{x{2.5cm} x{3cm} x{3.5cm} x{2cm} x{2cm} x{2cm}}
      \toprule
      \textbf{Defect} & \textbf{Energy} &  \textbf{DFT/Experiment} & \textbf{XMEAM-V} & \textbf{DP-HYB-V} & \textbf{GAP-V}~\cite{byggmastar_2020_prm} \\
      \hline
      \multirow{2}{*}{monovacancy} & formation energy     &  2.34, 2.6~\cite{han_2003_jap} (2.61~\cite{ma_2019_3_prm}) / 2.1-2.1~\cite{ullmaier_1991_adm} &    2.55 (2.56)       &  2.40 (2.41)   &   2.56 (2.58)      \\
      & migration energy &  0.65~\cite{ma_2019_3_prm} / 0.5~\cite{ullmaier_1991_adm} & 0.56 & 0.56 & 0.39 \\
      \hline
      \(\langle 111 \rangle\) dumbbell & &  \textbf{2.71} (\textbf{2.91}~\cite{ma_2019_prm}) &  \textbf{2.83} (\textbf{2.87})               &  \textbf{2.83} (\textbf{2.93})  &   \textbf{2.76} (\textbf{2.89})          \\
      \(\langle 110 \rangle\) dumbbell & &  3.01 (3.16~\cite{ma_2019_prm}) &    2.98 (3.05)  &  us~\tnote{a} (us~\tnote{a} )  &   us~\tnote{a} (us~\tnote{a} )      \\
      \(\langle 100 \rangle\) dumbbell & formation &  3.20 (3.38~\cite{ma_2019_prm}) &    3.21 (3.25)        &  us~\tnote{a} (3.85)     &   us~\tnote{a} (3.42)      \\
      \(\langle 111 \rangle\) crowdion   & energy  &  2.71 (\textbf{2.91}~\cite{ma_2019_prm}) &    us~\tnote{a} (\textbf{2.87})  &  us~\tnote{a} (2.96)  &   us~\tnote{a} (\textbf{2.89})     \\
      tetrahedral  &  &  3.23 (3.42~\cite{ma_2019_prm}) &    us~\tnote{b} (3.34)   &  us~\tnote{a} (us~\tnote{a} )  &   us~\tnote{a} (us~\tnote{a} )      \\
      octahedral   &  &  3.27 (3.44~\cite{ma_2019_prm}) &    us~\tnote{b} (3.35)     &   us~\tnote{a} (3.84)    &   us~\tnote{a} (3.49)     \\
     \bottomrule
    \end{tabular}
    \begin{tablenotes}
    \item[a] The structure relaxes to the \(\langle 111 \rangle\) dumbbell configuration.
    \item[b] The structure relaxes to the \(\langle 110 \rangle\) dumbbell configuration.
    \end{tablenotes}
  \end{threeparttable}

\end{table*}

\begin{figure*}[!htbp]
  \centering
  \includegraphics[width=1.0\textwidth]{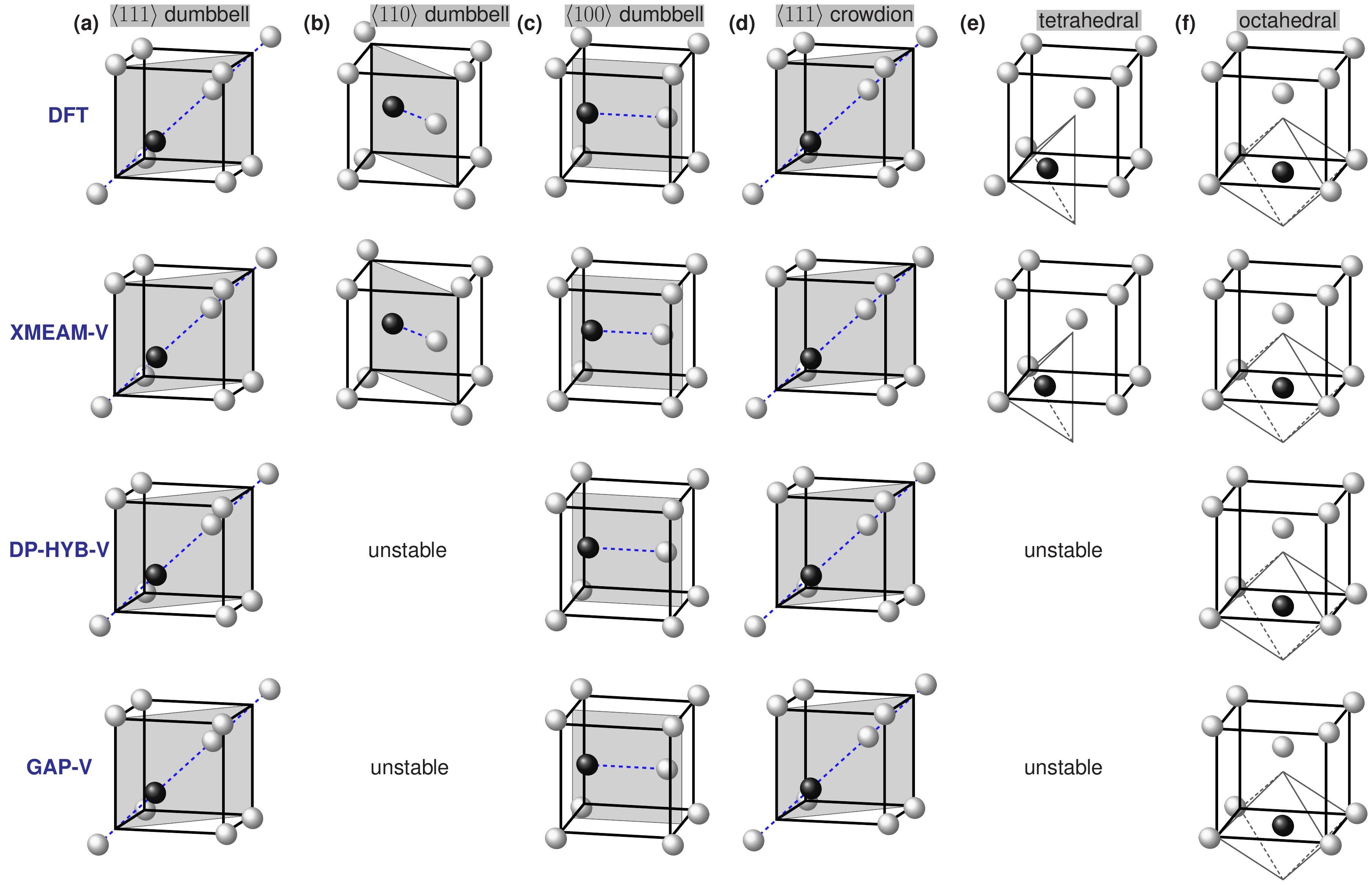}
  \caption{\label{fig:self_interstitial} Self-interstitials in the BCC structure of V predicted by DFT, XMEAM-V, DP-HYB-V and GAP-V. (a) \(\langle 111 \rangle\) dumbbell. (b) \(\langle 100 \rangle\) dumbbell. (c) \(\langle 110 \rangle\) dumbbell.  (d) \(\langle 111\rangle\) crowdion. (e) tetrahedral. (f) octahedral. All structures are the optimised configurations under the fixed supercell constraint. All models have similar self-interstitial structures while quantitative differences still exist in atom positions. The self-interstitial atom is shown in dark color.}
\end{figure*}

BCC TMs are often employed in high temperature and irradiative environments.  Point defects can be generated frequently, and accumulate to high densities and form various defect clusters, which in turn directly affect their macroscopic mechanical properties.  We consider the monovacancy and 6 self-interstitials in BCC V : \(\langle 111 \rangle\) dumbbell, \(\langle 110 \rangle\) dumbbell, \(\langle 100 \rangle\) dumbbell, \(\langle 111 \rangle\) crowdion, tetrahedral, and octahedral interstitials (Fig.~\ref{fig:self_interstitial}).  These defects can interact with themselves, dislocations, grain/interface boundaries, and crack-tips.  Table~\ref{tab:v_point_defect} summarizes the point defect energies of DFT, XMEAM-V, DP-HYB-V, GAP-V and experiments. In DFT, all formation energies with the OSC method are generally \(\sim\)0.2 eV lower than previous DFT values with the FSC method~\cite{ma_2019_prm}.  These energy differences may also arise from the different DFT parameters (cutoff energy, \(k\)-point densities) employed.  However, the lower formation energy is expected in the OSC method, and is also seen in calculations with the interatomic potentials.

For the monovacancy, all three potentials give formation and migration energies comparable to DFT and experimental values.  For self-interstitials, all configurations are at least meta-stable in both the OSC and FSC methods in DFT, while some relax to lower energy configurations in calculations using the interatomic potentials (Fig.~\ref{fig:self_interstitial}).  The \(\langle 111 \rangle\) dumbbell is predicted as the ground state configuration by DFT and all the three potentials, followed by the \(\langle 111 \rangle\) crowdion configuration at 5 meV higher in the OSC method in DFT.  The \(\langle 111 \rangle\) crowdion is not stable and relax to the \(\langle 111 \rangle\) dumbbell using the interatomic potentials and the OSC method.  The \(\langle 110 \rangle\) and \(\langle 100 \rangle\) dumbbell configurations have the third and fourth highest energies and are meta-stable in XMEAM-V, while DP-HYB-V and GAP-V predict that the \(\langle 110 \rangle\) and \(\langle 100 \rangle\) dumbbells are not stable using the OSC method and show only the \(\langle 100 \rangle\) dumbbell is meta-stable in the FSC method. All other self-interstitials are not stable using the OSC method and the interatomic potentials.  For the stable/metastable configurations, quantitative differences on atomic positions at point defects exist among the different computational models.  Nevertheless, all potentials generally reproduce the stable and meta-stable self-interstitial energetics and structures in good agreement with DFT (Fig.~\ref{fig:self_interstitial}).

\subsection{Phonon and temperature-dependent lattice properties}
\begin{figure}[!htbp]
  \centering
  \includegraphics[width=0.45\textwidth]{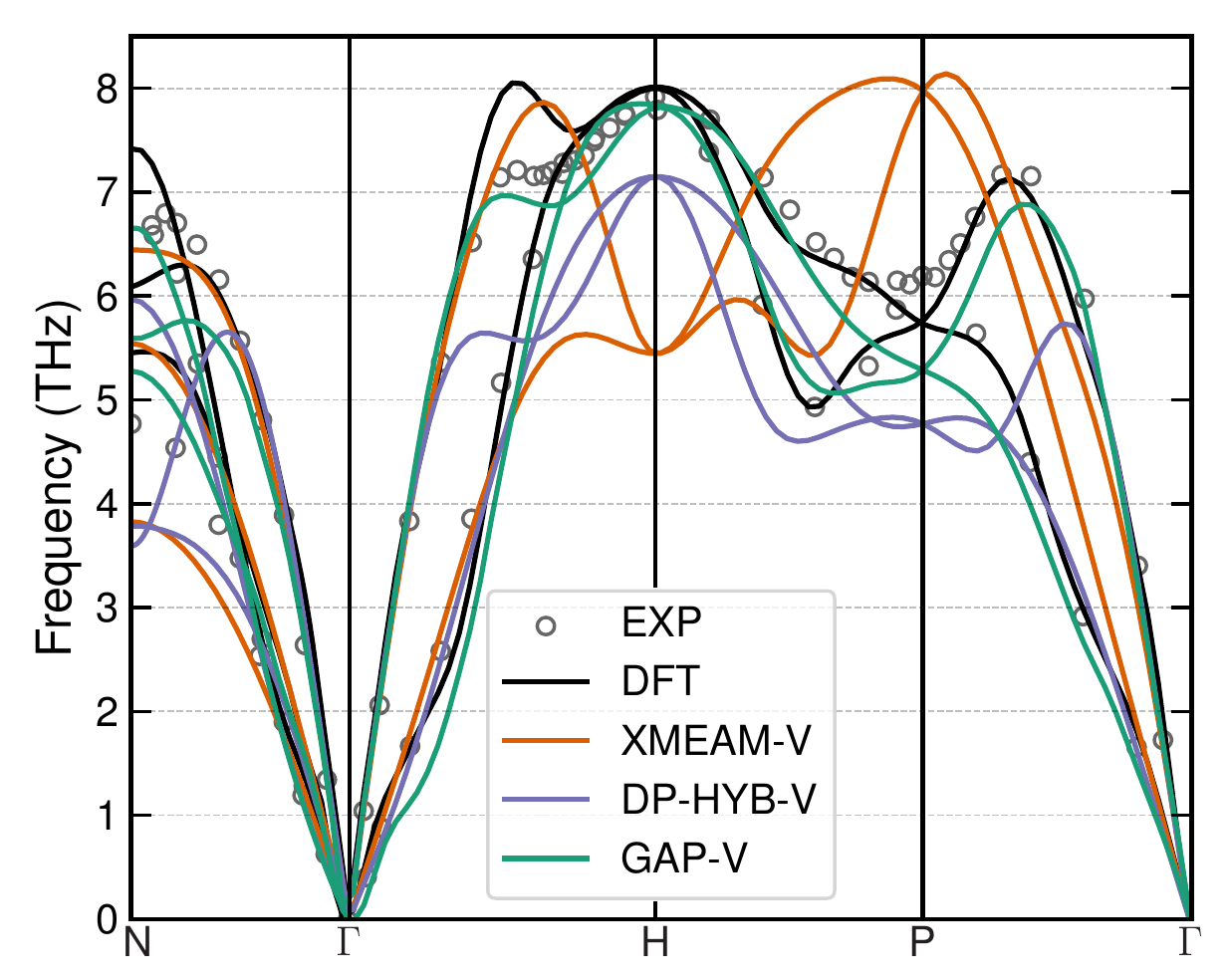}
  \caption{\label{fig:phonon} The phonon spectra of BCC phase from DFT, XMEAM, GAP, DP, and experiment. The DFT and experimental results are from Refs.~\cite{luo_2007_pnas,bosak_2008_prb}.}
\end{figure}

Figure~\ref{fig:phonon} shows the phonon spectra of BCC V at 0 K based on XMEAM-V, DP-HYB-V, GAP-V, DFT~\cite{luo_2007_pnas} and experiments~\cite{bosak_2008_prb}.  At low to medium frequencies, all interatomic potentials and DFT agree well with experimental data.  In particular, XMEAM-V has slightly better agreement with experiment around the \(\Gamma\) point, which reflects its accurate elastic constants of the BCC structure.  However, at higher frequencies in the \(N\), \(H\), and \(P\) directions, the two ML potentials are more accurate than XMEAM-V. The XMEAM-V potential does not qualitatively reproduce the basic symmetry of the phonon spectrum at the \(H\)- and \(P\)-points.  GAP-V is particularly close to the DFT and experiment data, and is expected to be more accurate in reproducing properties such as thermal conductivity and diffusivity, while XMEAM-V is perhaps more accurate in describing mechanical properties such as crack-tip dislocation nucleation given its promising shear modulus in a wide range of temperatures (see below).

\begin{figure*}[!htbp]
  \centering
  \includegraphics[width=0.95\textwidth]{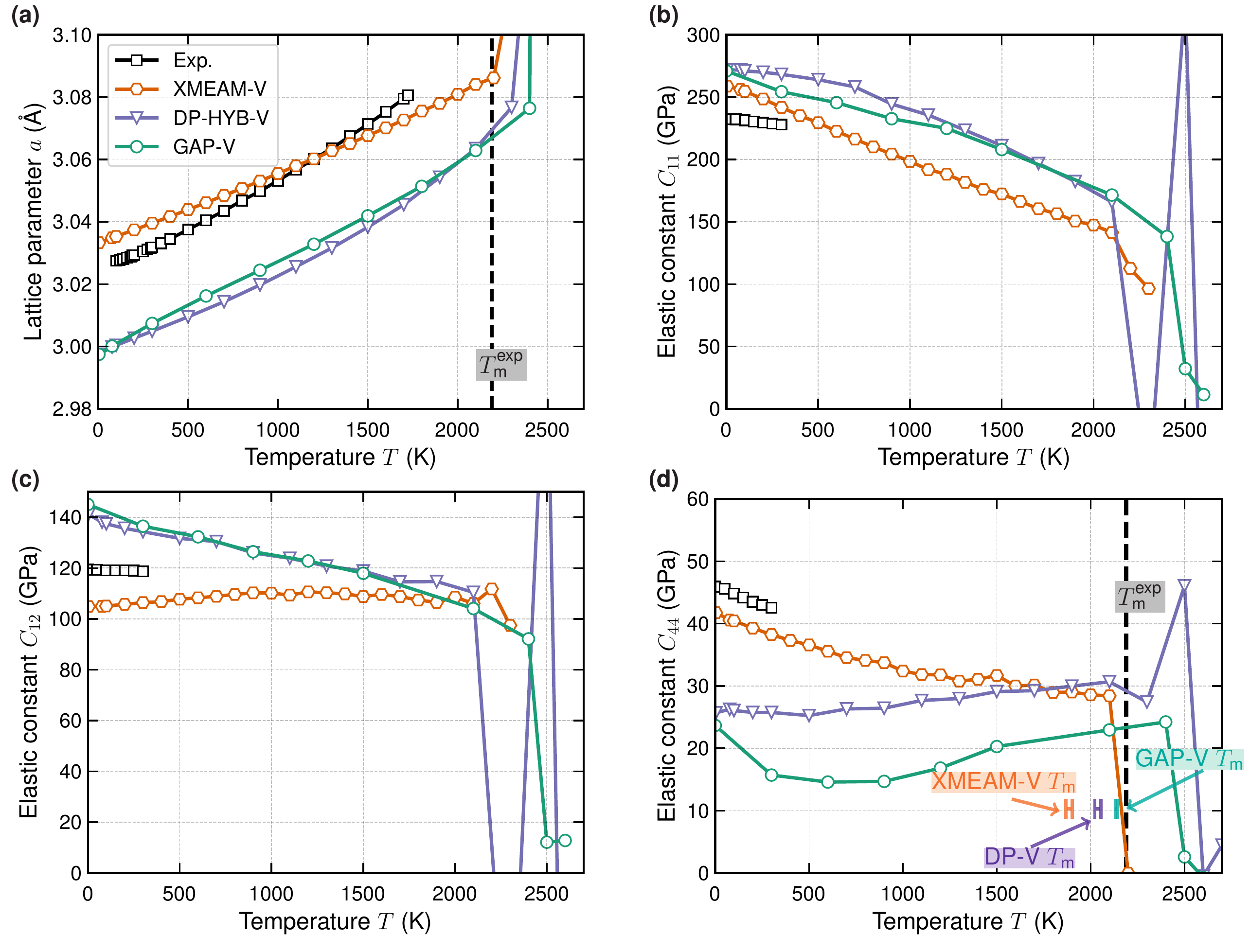}
  \caption{\label{fig:lat_els_t} Lattice parameter \(a\) and elastic constants of BCC V at finite temperatures. The experimental data of lattice parameters and elastic constants are from Refs.~\cite{kozlovskii_2020_jpcs} and~\cite{alers_1960_pr}, respectively. Dashed black lines in (a) and (d) mark the experimental melting temperature of V at 2183K~\cite{rumble_2019_crc}. The melting temperatures obtained from the solid-liquid two phase method are shown in (d).}
\end{figure*}

Figure~\ref{fig:lat_els_t} shows the BCC lattice parameter \(a(T)\) and elastic constants \(C_{ij}(T)\) as a function of temperature calculated by the three interatomic potentials and from experiments~\cite{alers_1960_pr,kozlovskii_2020_jpcs}.  All three potentials yield variations of lattice parameter with temperature in excellent agreement with experiment over the entire temperature range. In particular, XMEAM-V has accurate lattice parameters with discrepancies less than 0.01 {\AA} while the two ML potentials underestimate the lattice parameter by \(\sim\)0.03 {\AA}.  The offsets of \(a(T)\) of the ML potentials are likely inherited from DFT which exhibits a similar offset at 0 K.  The linear coefficient of thermal expansion (\(\alpha(T) = 1/a(\partial a /\partial T)\)) are \(7.02\times 10^{-6}\text{ K}^{-1}\), \(7.49\times 10^{-6}\text{ K}^{-1}\), and \(10.31\times 10^{-6}\text{ K}^{-1}\) for XMEAM-V, DP-HYB-V, and GAP-V at 300 K respectively, which are close to the experimental value of \( 8.71\times 10^{-6}\text{ K}^{-1}\). Near 2200-2300 K, the slope of \(a(T)\) changes in all potentials, indicating a phase transition at this temperature.

For the finite-temperature elastic constants, GAP-V and DP-HYB-V exhibit continuous decrease of \(C_{11}(T)\) and \(C_{12}(T)\) in the entire temperature range, while XMEAM-V has \(C_{11}(T)\) decreasing continuously up to 2100 K and \(C_{12}(T)\) almost independent of temperature.  The temperature insensitivity of \(C_{12}(T)\) agrees well with the experimental data up to 300 K. For the shear modulus \(C_{44}(T)\), XMEAM-V agrees well with experiments in both its magnitude and slope. DP-HYB-V has \(C_{44}(T)\) nearly independent of \(T\) at low temperatures and then increasing slowly with \(T\) at high temperatures, while \(C_{44}(T)\) of GAP-V decreases rapidly at low temperatures and gradually increases with increasing temperatures.  We suspect that the pathologies in \(C_{44}(T)\) of the ML potentials are associated with the inaccuracies of DFT for the shear constant of V at \(T=0\).  Nevertheless, their slopes of \(C_{44}(T)\) are unusual with respect to experimental results. In the intermediate temperature range, discrepancies of the shear modulus \(C_{44}(T)\) is further enlarged in GAP-V.

All of the elastic constants drop precipitously at a temperature close to the experimental solid-liquid transition temperature.  In particular, the shear modulus \(C_{44}(T)\) of XMEAM-V drops to 0 at 2200 K, suggesting the BCC phase transforms to the liquid phase, in agreement with the experimental melting temperature of 2183 K~\cite{rumble_2019_crc}.  The two ML potentials have their \(C_{44}(T)\) reaching 0 at about 2500 K.  The shear modulus data \(C_{44}(T)\) give some approximations of the melting temperatures of the respective potentials.  In the solid-liquid coexistence method, the obtained melting temperatures are consistently lower than these approximations (Fig.~\ref{fig:lat_els_t}d).  Specifically, the melting temperatures of XMEAM-V and DP-HYB-V are determined to be \(1875 \pm 25\) K and \(2025 \pm 25\) K, respectively, while GAP-V has \(T_\text{m}\) reported as 2130 K~\cite{byggmastar_2020_prm}.  Melting temperatures from \(C_{44}\) are the upper limits, since the configurations used are ideal BCC structures in fully periodic supercells and homogeneous nucleation of liquid phases requires some barriers to be overcome and occurs at some higher temperatures.  Nevertheless, all potentials have melting temperature around 2000 K close to experimental values.

Overall, XMEAM-V has better lattice and elastic properties at finite temperatures than the two ML potentials, despite the fact that finite-temperature properties are not included explicitly in the fitting of XMEAM-V.  For the two ML potentials (Table~\ref{tab:dp_training_data} and Ref.~\cite{byggmastar_2020_prm}),  the training datasets include AIMD configurations and yet their \(C_{44}\) values vary considerably in the intermediate temperature range and rise at high temperatures, indicating that additional datasets may be needed to reproduce the shear modulus at finite temperatures in ML frameworks.

\subsection{Dislocation core structures}

\begin{figure*}[!htbp]
  \centering
  \includegraphics[width=0.75\textwidth]{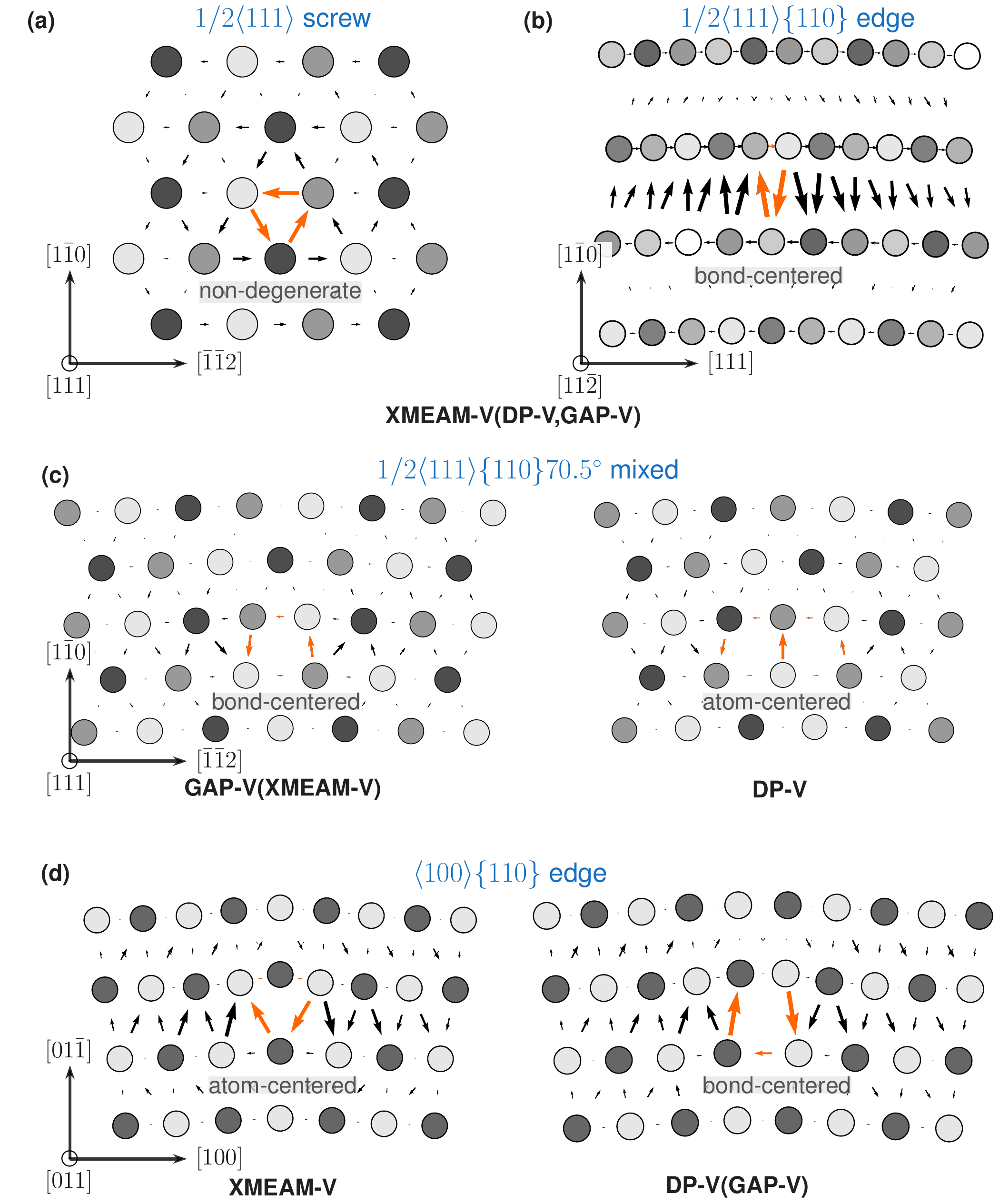}
  \caption{\label{fig:core} Core structures of \(\langle111\rangle/2\) and \(\langle 100 \rangle\) dislocations in BCC V.  (a-b) The non-degenerate \(\langle111\rangle/2\) screw core and the \(\langle111\rangle/2\) edge core on the \(\{110\}\) plane predicted by XMEAM-V, DP-HYB-V and GAP-V. (c) The bond-centered and atom-centered \(1/2\langle 111 \rangle\{110\}\) 70.5\(^{\circ}\) mixed core on the \(\{110\}\) plane predicted by XMEAM-V/GAP-V and DP-HYB-V, respectively.  (d) The \(\langle {100} \rangle \{110\}\) edge core on the \(\{110\}\) plane predicted by XMEAM-V and GAP-V/DP-HYB-V.  All cores are visualized with the differential displacement map.  For the edge and mixed cores, edge components are plotted on the relaxed configurations. For the screw core, screw components are plotted on the ideal BCC lattice.}%
\end{figure*}

Dislocations are the primary plastic strain carriers in most metals at low to moderate temperatures. In BCC V, plastic slip occurs via the motion of dislocations with a \(\langle 111 \rangle/2\) Burgers vector.   In BCC TMs, the \(\langle 111 \rangle/2\) screw dislocation has high lattice friction, carries most plastic strain and thus dictates stress responses at low temperatures.  The screw dislocation exhibits a wide range of intriguing behaviours which originate from its core structures and associated properties.  Given its importance, the screw dislocation has been extensively studied in simulations using DFT or interatomic potentials.  As mentioned in the Introduction, existing interatomic potentials of V have limited capabilities in reproducing the screw dislocation properties and current understanding is largely derived from DFT calculations which are often limited to small supercell sizes and 0 K temperature, in addition to its inaccuracy in the shear elastic modulus \(C_{44}\).

We examine the structures of 4 different dislocations in BCC V using XMEAM-V, DP-HYB-V and GAP-V and discuss the results with reference to earlier DFT calculations of core properties in BCC TMs.  Figure~\ref{fig:core} shows the core structures of the \(\langle 111 \rangle/2\) screw, edge, 70.5\(^\circ\) mixed and \(\langle 100 \rangle\) dislocation predicted by the interatomic potentials at 0 K. The \(\langle 100 \rangle\) edge dislocation was previously observed in BCC Mo~\cite{cheng_2013_jmr} while the mixed dislocation may exhibit high lattice friction as proposed in an earlier study~\cite{kang_2012_pnas}.  For the screw dislocation (Fig.~\ref{fig:core}a), all potentials produce the ND core, consistent with previous DFT calculations~\cite{clouet_2021_crp} and the prediction based on a new material index \(\chi\)~\cite{wang_2021}.  All potentials also predict a non-dissociated core of the \(\langle {111} \rangle/2\) edge dislocation on the \(\{110\}\) plane (Fig.~\ref{fig:core}b), consistent with the \(\gamma\)-surfaces of the \(\{110\}\) plane (Fig.~\ref{fig:pot_gamma_surface}) where no meta-stable stacking fault exists.

Figure~\ref{fig:core}c shows the core structures of the mixed dislocation. XMEAM-V and GAP-V produce a bond-centered (BC) structure while DP-HYB-V gives the atom-centered (AC) structure.  The BC core is seen in BCC TMs (Nb, Ta, Fe, Mo, W) in DFT calculations and the AC core is observed in some interatomic potentials~\cite{romaner_2021_am}.  Since DFT suggests that the BC core is prevalent in other BCC TMs, XMEAM-V and GAP-V are likely producing the correct ground state core structure of the mixed dislocation in V.  Finally,  for the \(\langle{100}\rangle\) edge dislocation on the \(\{110\}\) plane (Fig.~\ref{fig:core}d), XMEAM-V shows a compact core structure, similar to the core in Fe from DFT~\cite{fellinger_2018_prm}.  In contrast, both DP-HYB-V and GAP-V predict a relatively open structure, which is also seen in GAP-Fe.  Since DFT suggests Fe adopts the compact structure and GAP-Fe produced the open structure, it is likely that the open structure in GAP-V/DP-HYB-V is not the ground state core structure.  Based on all of the above cores and available DFT results, XMEAM-V perhaps correctly produces all the ground state core structures.

\subsection{Dislocation Peierls stresses}

We further calculate the Peierls stresses \(\tau_\text{P}\) of the dislocations shown in Fig~\ref{fig:core}.  Table~\ref{tab:v_dis_ps} shows the computed results and DFT/Experimental data.  For all the dislocations and in all models, the \(\langle 111 \rangle/2\) screw dislocation has the highest \(\tau_\text{P}\), in agreement with TEM study where screw dislocations have low mobilities and are often observed as long straight lines (e.g. Nb at 50 K~\cite{louchet_1975_am}, Ta~\cite{nasser_1998_am} and W~\cite{caillard_2018_w_am}).  Specifically, XMEAM predicts \(\tau_{\text{P-screw}}\) as 1217 MPa, in agreement with the DFT calculation of 1000-1200 MPa~\cite{dezerald_2014_prb}, while DP-HYB-V and GAP-V have higher \(\tau_{\text{P-screw}}\) of 1961 and 1971MPa, respectively. For the edge dislocation, all models predict low \(\tau_\text{P}\), negligible compared to that of the screw core.  For the 70.5\(^{\circ}\) mixed dislocation, XMEAM-V and GAP-V have the same BC core structure and similar \(\tau_\text{P}\) at 30 and 79 MPa, while DP-HYB-V adopts the AC core with a much higher \(\tau_\text{P}\).  Low \(\tau_\text{P}\) of non-screw cores by XMEAM-V and GAP-V is consistent with (i) recent DFT calculations~\cite{romaner_2021_am} where the mixed core has nearly zero Peierls barriers and stresss and (ii) internal friction (IF) experiments where the two low-temperature peaks in IF spectra almost coincide~\cite{romaner_2021_am} in Group VB elements (Nb and Ta)~\cite{schultz_1991_msea}.  DP-HYB-V thus likely over-estimates the Peierls stress of these non-screw dislocations, as in other earlier interatomic potentials~\cite{kang_2012_pnas}.  For the \(\langle 100 \rangle\) edge dislocation on the \(\{110\}\) plane, XMEAM-V predicts its Peierls stress of 198 MPa, while DP-HYB-V and GAP-V have high Peierls stresses at 1102 MPa and 503 MPa.  For the mixed and edge cores, the discrepancies among the interatomic potentials seem to lie in their different core structures (Fig.~\ref{fig:core}), which adds further complexities in modelling and understanding dislocation and plastic deformation in BCC TMs. All three interatomic potentials have exactly the same relative ordering of the Peierls stresses of all the dislocations. Nevertheless, based on the available DFT and experimental results in the BCC TM family, XMEAM-V appears the preferred choice for modelling core structures and Peierls stresses.

\begin{table*}[!htbp]
  \caption{\label{tab:v_dis_ps} Peierls stress \(\tau_\text{P}\) (MPa) of the screw, mixed and edge dislocations in BCC V.} \centering
  \begin{threeparttable}
    \begin{tabular}{x{2.5cm} x{3.5cm} x{3.9cm} x{3cm} x{3cm}}
      \toprule
      \textbf{Model} &  \(1/2\langle 111 \rangle \) screw & \(1/2\langle 111 \rangle \{110\} 70.5^\circ\) mixed  &\(1/2\langle 111 \rangle \{110\}\) edge   & \(\langle 100 \rangle \{110\}\) edge   \\
      \hline
      XMEAM-V     &  1217      &  30    &  8  & 198 \\
      DP-HYB-V        &  1961      &  620   &  57 & 1102 \\
      GAP-V~\cite{byggmastar_2020_prm}       &  1971      &  79    &  20 & 503 \\
      DFT  &  1000-1200~\cite{dezerald_2014_prb}   & - & - & - \\
      Experiment & 360~\cite{suzuki_1999_pma} & - & - & - \\
      \bottomrule
    \end{tabular}
  \end{threeparttable}
\end{table*}

\subsection{The \(\langle 111 \rangle/2\) screw core Peierls barriers and energetics}

\begin{figure*}[!htbp]
  \centering
  \includegraphics[width=0.95\textwidth]{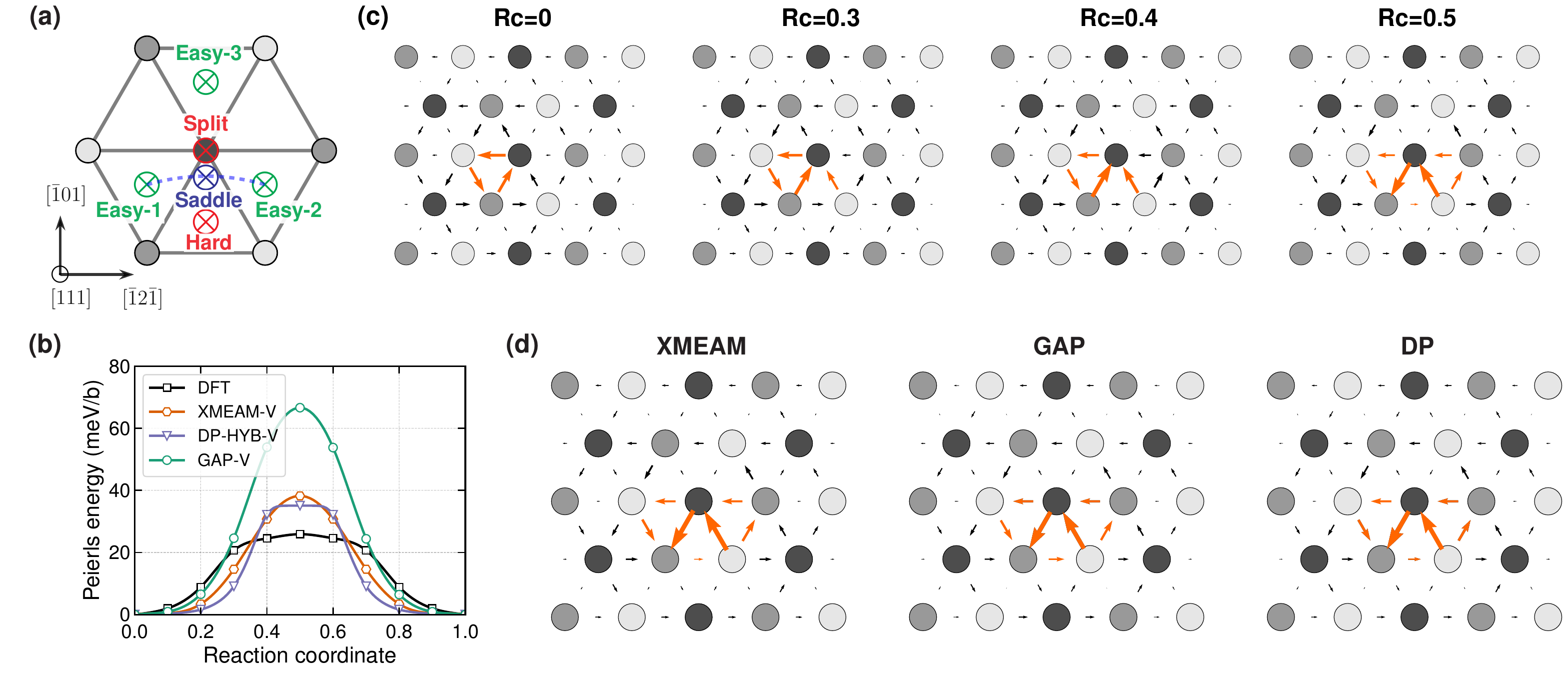}
  \caption{\label{fig:pp} Critical screw cores in 2D Peierls potential. (a) The schematic diagram of core positions (easy, hard, split and saddle cores) in the 2D Peierls potential. (b) Peierls energy of the screw dislocation calculated by the NEB method with interatomic potentials. DFT results are collected from Ref.~\cite{dezerald_2014_prb}. (c) Screw core migration process obtained from the NEB method with XMEAM-V. Rc denotes Reaction coordinate in the NEB calculations. Only the first half migration is shown as the process is symmetric. (d) Saddle core structures obtained by interatomic potentials.  Minor differences exist in the differential displacement (DD) between the two atoms below the core center.  All core structures are visualized with the DD map and core centers are highlighted with orange arrows.}
\end{figure*}

At finite temperatures, the glide of the \(\langle 111 \rangle/2\) screw dislocation is governed by the transition path and associated energy variation between two adjacent ground state core positions (Peierls valley).  This energy variation is known as the Peierls potential. We study the Peierls potential using the nudged elastic band (NEB) method.  Figure~\ref{fig:pp}a shows the schematics of the transition path and critical core positions viewed along the \(\langle 111 \rangle\) direction.  In particular, the easy, hard, split and saddle cores are highly related to the Peierls potential~\cite{dezerald_2014_prb}.  The easy and hard cores are at the centers of the triangles formed by three columns of atoms.  The relative positions of these three columns of atoms along the Burgers vector direction determine whether the structure is an easy or hard core. Previous DFT calculations have shown that the ND core always adopts the easy core position, while hard and split cores are the maximum energy states in BCC TMs~\cite{dezerald_2014_prb}.  The split core center is at the vicinity of one atomic column. It is often metastable in some EAM/MEAM potentials~\cite{mendelev_2007_prb,maisel_2017_prm}, which results in a double-hump Peierls energy profile.

Figure~\ref{fig:pp}b shows the Peierls energy profiles of XMEAM-V, DP-HYB-V and GAP-V calculated by the NEB method and previous DFT results~\cite{dezerald_2014_prb}.  In all the cases,  the energy profile is symmetric about the middle point of the transition path (Fig.~\ref{fig:pp}c) and has a single peak corresponding to the saddle core (Fig.~\ref{fig:pp}d) energy.  The Peierls barrier \(\Delta E_\text{PB}\) is thus determined by the energy difference between the saddle and easy cores and is the energy barrier per unit length that must be overcome for dislocation glide at 0 K.  All potentials exhibit similar saddle core structures with some minor differences on the magnitudes of the DD between the atom pair below the core center.  GAP-V predicts a Peierls barrier of 66.9 meV/\(b\),  close to that of W at 81.8 meV/\(b\)~\cite{dezerald_2014_prb} and much higher than the DFT values of 24.4 meV/\(b\)~\cite{weinberger_2013_prb} and 25.7 meV/\(b\)~\cite{dezerald_2014_prb}.  XMEAM-V and DP-HYB-V predict \(\Delta E_\text{PB}\) slightly above the DFT value, which itself may be underestimated in DFT (see below).  Table~\ref{tab:v_core_energy} summarizes the saddle, split and hard core energies relative to their respective easy cores predicted by the three potentials and DFT.  GAP-V has core energies at least 50\% higher than the corresponding DFT values, resulting in much higher Peierls barrier \(\Delta E_{\text{PB}}\).

\begin{table}[!htbp]
  \caption{\label{tab:v_core_energy} The energies of the saddle, split and hard cores relative to their respective easy cores predicted by DFT, XMEAM-V, DP-HYB-V and GAP-V. The energies are in the unit meV/\(b\). }
  \centering
  \begin{threeparttable}
    \begin{tabular}{x{1.6cm} x{2.5cm} x{2.0cm} x{2.0cm} }
      \toprule
      \textbf{Model} & \(\Delta E_{\text{saddle-easy}}\) & \(\Delta E_{\text{split-easy}}\) & \(\Delta E_{\text{hard-easy}}\)  \\
      \hline
      XMEAM-V & 38.2 & 56.4 & 74.5 \\
      DP-HYB-V & 35.1 & 67.6 & 77.3 \\
      GAP-V~\cite{byggmastar_2020_prm} & 66.9 & 136.0 & 108.7 \\
      DFT & 25.7~\cite{dezerald_2014_prb}, 24.4~\cite{weinberger_2013_prb} & 51.3~\cite{dezerald_2014_prb}  & 52.5~\cite{dezerald_2014_prb} \\
      \bottomrule
    \end{tabular}
  \end{threeparttable}
\end{table}

Previous DFT calculations show that the elastic constant \(C_{44}\) and Peierls barrier \(\Delta E_\text{PB}\) in V depend strongly on the number of valence electrons employed.  Specifically, \(C_{44}\) and \(\Delta E_{\text{PB}}\) are predicted to be 10.8 GPa and 14.6 meV/\(b\) with 5 valence electrons and increase to 22.0 GPa and 24.4 meV/\(b\) with 11 valence electrons~\cite{weinberger_2013_prb}.  As the experimental value of \(C_{44}\) is 46.0 GPa and nearly 2 times the DFT value, it is reasonable to expect the true \(\Delta E_\text{PB}\) is higher than the DFT value.  Linear extrapolations based on the shear modulus~\cite{weinberger_2013_prb} and valence electron number to all electrons will land \(\Delta E_{\text{PB}}\) at 51 meV/\(b\) and 44 meV/\(b\).  These extrapolations are not expected to be quantitatively accurate, or well founded since inner electrons have less influence than outer electrons. In any case, \(\Delta E_{\text{PB}}\) of XMEAM-V/DP-HYB-V are perhaps reasonable at this stage.

\subsection{Gliding of a long screw dislocation at finite temperatures}
\begin{figure*}[!htbp]
  \centering
  \includegraphics[width=0.95\textwidth]{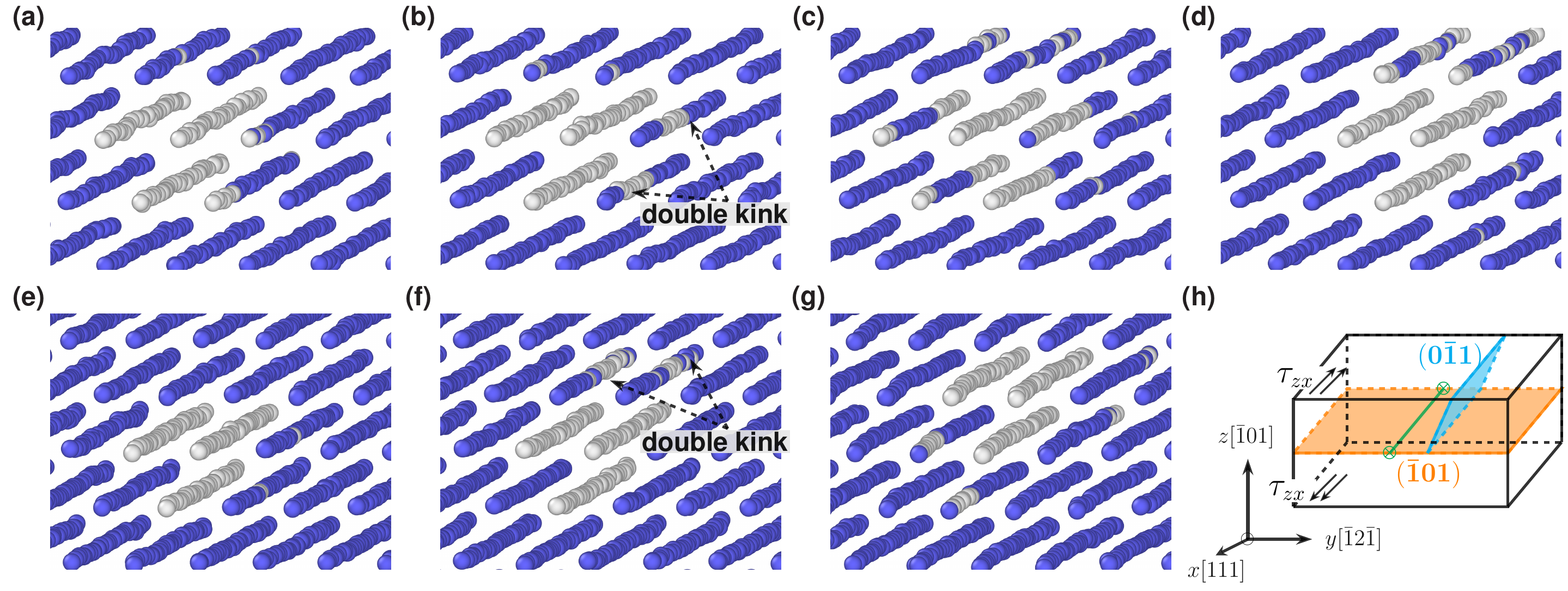}
  \caption{\label{fig:glide_77K_XMEAM} The gliding process of the long screw dislocation (30 \(\mathbf{b}\)) under a shear stress \(\tau_{zx}\) of 1 GPa at 77 K. Atoms are colored by their local structures identified by the common neighbor analysis: BCC-blue and white-others. (a-d) Double-kink nucleation and propagation along the \((\bar{1}01)\) MSSRP. (e-g) Double-kink nucleation and propagation along the lightly stressed \((0\bar{1}1)\) plane. (h) Schematics of the slip planes and the applied stress \(\tau_{zx}\).}
\end{figure*}

The screw dislocation is believed to glide via a double-kink nucleation and propagation mechanism at low to moderate temperatures.  We study its glide behaviour explicitly under a shear stress of 1 GPa at 77 K in MD simulations.  The applied stress is higher than the Peierls stresses in experiments (360 MPa~\cite{suzuki_1999_pma}), but enables quick examination of the elementary core migration step at relatively short timescales (e.g., 30 ps).  Since only the XMEAM-V gives accurate shear modulus, we focus on the finite temperature glide using XMEAM-V here.  Figure~\ref{fig:glide_77K_XMEAM} shows the atomic configuration during the core migration from one Peierls valley (easy core) to the next one (easy core) (a complete animation is available in the Supplementary Materials).  In the simulation, the screw core glides via the double-kink nucleation and migration mechanism.  In particular, the screw core is stationary at the easy core position for most of the time. A double kink of opposite signs is occationally nucleated at a short segment of the dislocation line on the maximum resolved shear stress plane (MRSSP, the \((\bar{1}01)\) horizontal plane in Fig.~\ref{fig:glide_77K_XMEAM}a-b), followed by the kink propagation in the opposite directions along the dislocation line.  The two kinks annihilate each other after they cross the periodic boundary and meet again, which completes the one-step migration of the entire screw dislocation.

In addition to the double-kink nucleation and migration along the MRSSP, a double-kink could occationally nucleate on the \((0\bar{1}1)\) plane with a lower resolved shear stress (Fig.~\ref{fig:glide_77K_XMEAM}e-g). The Schmid factor of the anomalous system \([111](0\bar{1}1)\) is only one half of that of the primary system \([111](\bar{1}01)\)~\cite{seeger_2002_pssa}.  This non-Schmid behaviour is surprising, but is consistent with the anomalous slip observed in high-purity V and more broadly in group VB and VIB TMs deformed at low temperatures~\cite{seeger_2002_pssa}.  In V in particular and with decreasing temperatures (e.g., 77 K), the tendency of slip on crystallographic \(\{110\}\) planes increases with frequent anomalous slips on lightly stresses \(\{110\}\) planes~\cite{taylor_1973_pm,creten_1977_mse,bressers_1977_sm} and branching on concurrent \(\{110\}\) planes~\cite{bressers_1982_jlcm}.  This anomalous slip is also widely observed in Nb~\cite{duesbery_1969_pm,bolton_1972_pm,louchet_1975_am,garratt_1976_pm,aono_1984_sm,groger_2018_jmr} and Ta~\cite{wasserbach_1995_pssa} and group VIB TM family (see Ref.~\cite{seeger_2002_pssa}).  XMEAM-V thus demonstrates its capability in modeling fundamental dislocation glide behaviour for BCC V, including the unexpected anomalous slip which seems to be an intrinsic property of the screw core.

\subsection{Computational speed}

\begin{figure}[!htbp]
  \centering
  \includegraphics[width=0.45\textwidth]{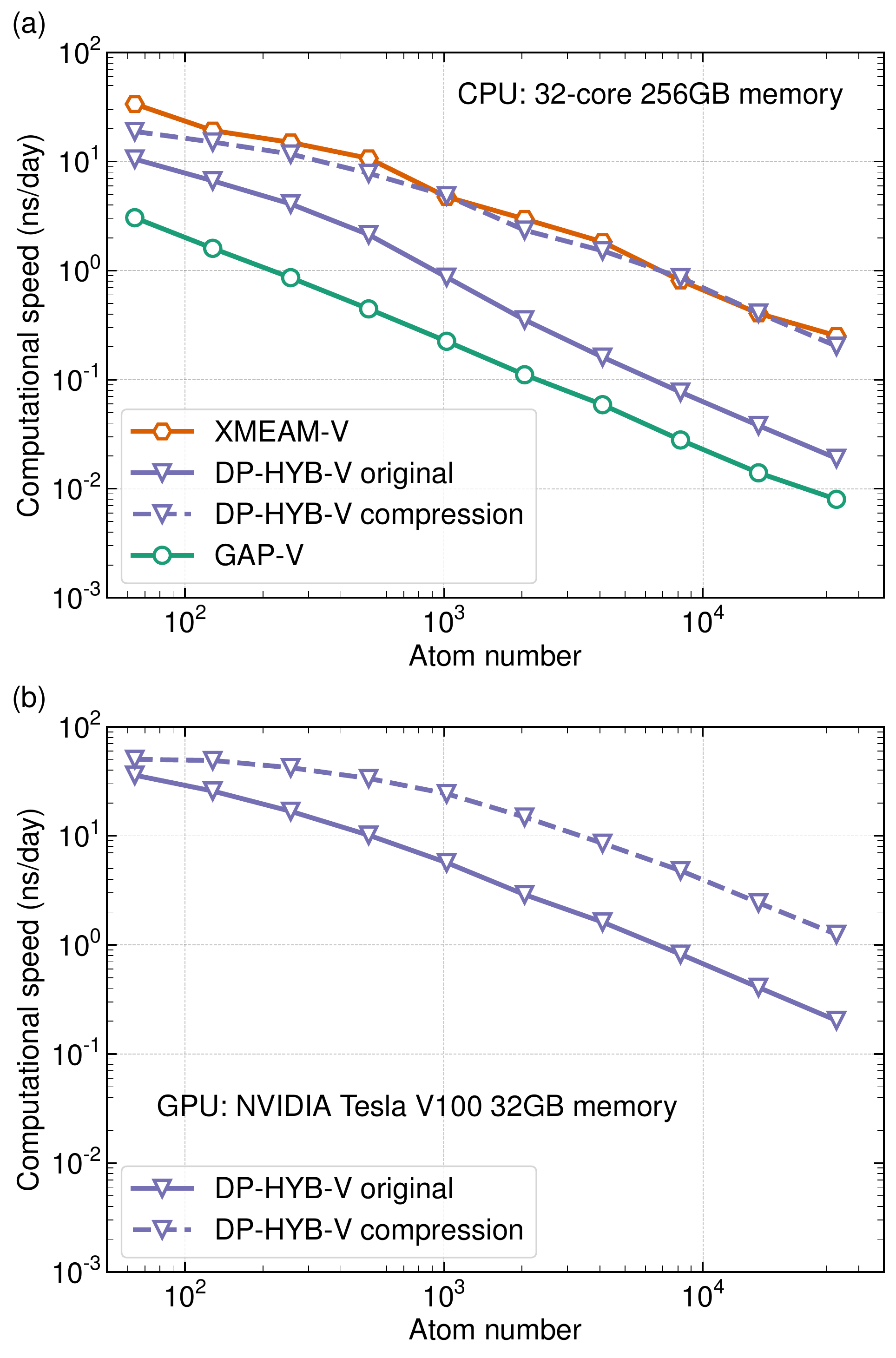}
  \caption{\label{fig:pot_speed} Speed comparison of the XMEAM-V, GAP-V, DP-HYP-V original, and DP-HYB-V compression model. (a) On a CPU machine. (b) DP-HYB-V original and compression models on a GPU machine. The computational speed (ns/day) is directly read from the log files of LAMMPS.}
\end{figure}

Finally, we compare the computational speeds of XMEAM-V, DP-HYB-V and GAP-V. All benchmarks are performed through the LAMMPS interfaces on a 32-CPU-core node and on a V100 GPU (only for DP). DP-HYB-V supports two models: the original and compressed models. The former preserves the exact information of the embedding neural network while the latter accelerates the computational speed via tabulating the embedding network~\cite{lu_2022_jctc}.  The benchmark measures the time elapsed for 10000 MD timesteps for supercells of perfect BCC structure at 100 K.  Figure~\ref{fig:pot_speed} shows the speed (ns/day) as a function of total number of atoms in the supercells.  Overall, all interatomic potentials exhibit near-linear scaling with increasing number of atoms, which is of utmost importance and a key attribute in stark contrast with first-principles DFT calculations.  For the largest system with more than 30000 atoms, XMEAM-V is slightly faster (20\%) than DP-HYB-V compression model, 13.3 times of DP-HYB-V original model, and 31.6 times of GAP-V on CPU.  The compression model accelerates DP-HYB-V by about a factor of 10 on CPU and 6 on GPU.  The above benchmark provides a general comparison among the different interatomic potential formalisms and are only for references to estimate their respective computing costs.  Further code optimizations are certainly possible in the various models.

\section{Discussion}
\label{sec:discussion}

Interatomic potentials provide a bridge between material properties at lattice scales and defect properties at micromechanics scales.  This upscaling is achieved by approximating the potential energy landscape of the system with analytical, numerical/spline-curve, Gaussian basis, or neural-network functions. GAP does not have fixed functional form and can be systematically improved, so does DP which employs variable-size embedding and fitting neural-network functions.  Both GAP and DP have well-established training frameworks~\cite{byggmastar_2020_prm,wang_2018_cpc} which allows systematic development of new interatomic potentials.  For V, both DP-HYB-V and GAP-V are trained with a broad range of datasets generated from first-principles calculations.  GAP-V is trained with BCC structures (elastically distorted, high temperatures, vacancy, self-interstitials, surface, \(\gamma\)-surfaces), elastically distorted FCC, HCP, simple cubic, diamond, A15 and C15, as well as liquids and dimers.  DP-HYB-V is trained with a similar datasets covering a smaller range of structures (mainly BCC and FCC).  Both these ML-type potentials accurately reproduce V properties from DFT.  In particular, both potentials possess the ND core structure and single-hump Peierls energy profile of the \(\langle 111 \rangle/2\) screw dislocation, which has been a challenging task for interatomic potentials over several decades.  This success, however, has its origin in the crystal geometry at the screw core center.  We recently discovered that the screw dislocation core structure is governed by the cohesive energy difference \(\Delta E_\text{FCC-BCC}\) bewteen the FCC and BCC structures~\cite{wang_2021}.  Since DP-HYB-V and GAP-V are fit to both BCC and FCC structures and reproduce their cohesive energies accurately (Table~\ref{tab:basic_prop}), they naturally produce the ND core.

The capability of DP and GAP to fit to multi-structure/state is an inherent advantage in ML type of interatomic potentials, since they use extensible functions and allow continuous improvements.  Training with multi-structure has further profound impacts on the transferability of interatomic potentials in general.  For example, recent DFT calculations show that the \(\langle 111 \rangle/2\) screw dislocation Peierls barrier and nucleation barrier scale linearly with \(\Delta E_\text{FCC-BCC}\) in BCC TMs~\cite{wang_2021}.  Previous first-principles calculations show that the tetragonal shear constant \(C^{\prime}=(C_{11}-C_{12})/2\) is also determined by \(\Delta E_\text{FCC-BCC}\) and the Bain path~\cite{soderlind_1993_prb}.  While all these are ultimately related to band-filling or valance electron concentrations~\cite{wang_2021}, these interatomic potentials do not contain electrons explicitly.  Therefore, it is likely that electronic structures/quantum mechanics information is transferred to ML potentials more robustly via structures of different phases.  Nevertheless, such transfer is not always straightforward or automatic in ML frameworks, nor guarantees accurate reproduction of any particular properties.  For example, GAP-V has its screw dislocation saddle, split and hard core energies 100\% higher than the corresponding DFT values, while the Peierls potential can only be accurately reproduced with the new DP-HYB containing a three-body descriptor in the DP framework, despite their accuracies with respect to DFT in many other properties including the unstable FCC phase and the erroneous elastic constant \(C_{44}\).  Of course, such ML potentials can be no better than the datasets upon which they are trained.  In the present case, the datasets come from DFT in the GGA approximation.  At this level of electronic structure calculations, it is well-known that DFT does poorly for Group VB elements and especially so for V.  This has variously been attributed to orbital localization and delocalization error in elements with strongly localized and correlated valence electrons~\cite{wang_2020_jcp}.  Hence, Group VB V should be viewed as a worst case scenario.  The resulting problem is poor reproduction of the  elastic constant \(C_{44}\).  Recent semi-empirical DFT+J formalisms correct such a problem~\cite{wang_2020_jcp}, leading to more accurate \(C_{44}\) of V.  Hence, the accuracies of ML potentials may be improved by retraining the entire potential using the DFT+J approach.

MEAM/XMEAM uses the classical formalism with analytical functions for its electron density and embedding functions, and a spline curve for its pair-interaction function.  It is semi-empirical with its total energy expression conceptually related to the tight-binding theory, has fixed functional form and contains 18/24 fitting parameters. The MEAM/XMEAM also has the flexibility to be trained using any combination of datasets from DFT and experiments.  Compared with the ML potentials,  the classical semi-empirical MEAM/XMEAM contains considerably fewer parameters and can be trained on small datasets compared with those used for ML potentials.  Nevertheless, the selection of the datasets requires \textit{a priori} knowledge about the material system (ML frameworks also require some of it).  The original MEAM faces considerable difficulty in reproducing properties of multiple structures, which is crucial for many defect properties and phase transformations.  The new XMEAM extends the MEAM capability for multi-structure, as seen in earlier similar works~\cite{baskes_2007_prb,gibson_2017_msmse}.  Nevertheless, it also has limitations.  For example, the current XMEAM-V cannot reproduce the \(\Delta E_\text{FCC-BCC}\) of V at the DFT value; raising \(\Delta E_\text{FCC-BCC}\) leads to deterioration of other properties.  However, preliminary results of an XMEAM potential for W suggest it is possible to reproduce \(\Delta E_\text{FCC-BCC}\) of W with reasonable overall properties.

While XMEAM-V is the first potential developed under this extended MEAM framework, further refinements are possible, such as through the selection of fitting datasets, procedures, parameter space and weights.  ML potential frameworks, such as the GAP and DP, are also evolving towards more accurate material and alloy properties and increased efficiency (e.g., DP on GPUs and tabulated GAP~\cite{byggmastar_2022_prm}).  However, their accuracy for group VB TMs will continue to be limited by the current functionals in DFT calculations until a practical improvement/replacement is found and validated (e.g., DFT+J).  Nevertheless, the improvements in DFT, at almost no additional computational cost, is tempting for ML potentials given their natural flexibility.  A broad survey on existing interatomic potentials show that very few interatomic potentials (include ML ones) can simultaneously give the correct ND screw dislocation core structure and accurate Peierls energy profile.  The issues shown in Table~\ref{tab:ex_pots} are thus not unique for V but general for all BCC TMs.  Given the current state-of-the-art, XMEAM offers a reasonable path applicable to all nonmagnetic BCC TMs.  The datasets and fitting procedures introduced here can be easily applied to Nb and Ta, and perhaps to Mo and W as well.  Compared to the current DP-HYB-V and GAP-V, XMEAM-V reproduces nearly all properties relevant to point defect, dislocation and fracture properties.  It offers a balance between accuracy and efficiency and thus can be used broadly to study plastic deformation at different loading conditions, temperatures, and likely with high fidelity.  In particular, the anomalous slip, dislocation mobility as a function of dislocation character (edge, mixed, screw), twinning, crack-tip behaviours should be examined more carefully than the simulations in the current work with a main purpose of introducing the new potentials.  Such specific studies will provide new insights and guidance on realistic plasticity modellings at higher scales such as in dislocation dynamics and crystal plasticity finite element analysis.

\section{Conclusion}
\label{sec:conclusion}
In summary, we developed two new interatomic potentials for BCC V, using an extended form of the classical, semi-empirical MEAM (XMEAM) and the machine-learning DP-HYB framework.  Both new potentials exhibit superior accuracy for mechanical properties relative to all existing interatomic potentials.  We performed comprehensive comparisons among the XMEAM and two ML potentials (DP and GAP) on the thermodynamic and mechanical properties of V.  The two ML potentials inherit the erroneous properties of V from current DFT calculations. On the other hand, the classical XMEAM potential, trained using a selection of experimental and DFT data, gives accurate properties relevant for plastic and fracture phenomena at both 0 K and finite temperatures.  In particular, XMEAM-V reproduces all screw, edge, mixed dislocation core structures, Peierls stress at 0 K and anomalous slip at 77 K, enabling large-scale atomistic simulations in BCC V.  XMEAM expands the capability of classical potentials for multi-structure and provides a practical path to developing interatomic potentials for other BCC TMs, and in particular the group VB TMs where the most widely used DFT functionals have limited accuracy.  Since XMEAM retains the essential features of MEAM, XMEAM interatomic potentials fit for pure elements may be used as a foundation for developing potentials for multi-principal element alloys, particularly the refractory class NbTaMoW alloys and its derivatives.

\section{Acknowledgments}

The work of R.W., X.M., and Z.W. is supported by the Research Grants Council (RGC), Hong Kong SAR through the Early Career Scheme (ECS) Fund under project number 21205019 and Collaborative Research Fund (CRF) under project number 8730054. The work of D.J.S. and T.W. is supported by RGC through CRF project 8730054. Computational resources are provided by the Computing Services Center, City University of Hong Kong.

\section{Appendix}

Table~\ref{tab:bcctm_elas_exp_dft_pot} shows the elastic constants predicted by DFT and machine learning interatomic potentials trained based on the DFT-computed datasets.  These values are used to compute the relative errors shown in Fig.~\ref{fig:elastic_c_tm}.

\begin{table*}[!htbp]
  \caption{\label{tab:bcctm_elas_exp_dft_pot} The elastic constants of BCC nonmagnetic transition metals. The DFT-1~\cite{byggmastar_2020_prm} and DFT-2~\cite{li_2020_npjcm} values are used to train the GAP~\cite{byggmastar_2020_prm} and SNAP~\cite{li_2020_npjcm} potentials, respectively. The experimental values are extrapolated to 0 K from a series of measurements at low temperatures~\cite{simmons_1971_mit}.}
  \centering
  \begin{threeparttable}
    \begin{tabular}{x{1.5cm} x{3.5cm} x{2.0cm} x{1.5cm} x{1.5cm} x{1.5cm} x{1.5cm}}
      \toprule
      \textbf{Element}  & \textbf{Elastic constants} (GPa)     & \textbf{Experiment}  & \textbf{DFT-1}~\cite{byggmastar_2020_prm}  & \textbf{GAP}~\cite{byggmastar_2020_prm}   & \textbf{DFT-2}~\cite{li_2020_npjcm} & \textbf{SNAP}~\cite{li_2020_npjcm} \\
      \hline
               & \(C_{11}\)        &  232.4 (0 K)      & 269    & 271   & -      & -     \\
      V        & \(C_{22}\)        &  119.4      & 146    & 145   & -      & -     \\
               & \(C_{44}\)        &  46.0       & 22     & 23    & -      & -     \\
      \hline
               & \(C_{11}\)        & 252.7 (4.2 K)& 237    & 243   & 249    & 266   \\
      Nb       & \(C_{22}\)        &  133.2      & 138    & 137   & 135    & 142   \\
               & \(C_{44}\)        &  30.8       & 11     & 13    & 19     & 20    \\
      \hline
               & \(C_{11}\)        &  266.3 (0 K) & 266    & 267   & 264    & 257   \\
      Ta       & \(C_{22}\)        &  158.2      & 161    & 161   & 161    & 161   \\
               & \(C_{44}\)        &  87.4       & 77     & 77    & 74     & 67    \\

      \hline
               & \(C_{11}\)        &  450.0 (0 K) & 468    & 472   & 472    & 435   \\
      Mo       & \(C_{22}\)        &  173.0      & 155    & 163   & 158    & 169   \\
               & \(C_{44}\)        &  125.0      & 100    & 105   & 106    & 96    \\

      \hline
               & \(C_{11}\)        &  532.6 (0 K) & 521    & 524   & 511    & 560   \\
      W        & \(C_{22}\)        &  205.0      & 195    & 200   & 200    & 218   \\
               & \(C_{44}\)        &  163.1      & 147    & 148   & 142    & 154   \\
      \bottomrule
    \end{tabular}
  \end{threeparttable}
\end{table*}

\section{References}

\end{document}